\documentclass[fleqn,usenatbib]{mnras}

\usepackage[T1]{fontenc}

\DeclareRobustCommand{\VAN}[3]{#2}
\let\VANthebibliography\thebibliography
\def\thebibliography{\DeclareRobustCommand{\VAN}[3]{##3}\VANthebibliography}

\usepackage{graphicx}	
\usepackage{amsmath}	
\usepackage{comment}
\usepackage{txfonts}
\usepackage{hyperref}
\usepackage{textgreek}
\usepackage{xargs}
\usepackage{xspace}
\usepackage{booktabs} 
\usepackage[dvipsnames,hyperref]{xcolor}

\usepackage{bigdelim,multirow}
\usepackage{paralist}
\usepackage{textgreek}
\usepackage{xargs}
\usepackage{xspace}
\usepackage{makecell}
\newcommand{\subref}[2]{\hyperref[#1]{\ref{#1}{#2}}}
\newcommand{\subreflett}[2]{\hyperref[#1]{#2}}

\usepackage{hyperref}
\hypersetup{
    colorlinks=true,
    linkcolor=blue,
    citecolor=blue,
    filecolor=magenta,      
    urlcolor=blue,
    }

\graphicspath{{./}{figures/}}
\usepackage{etoolbox}
\makeatletter
\newcommand\sendemail[3]{
\edef\@tempa{mailto:#1?subject=#2 }%
\edef\@tempb{\expandafter\html@spaces\@tempa\@empty}%
\href{\@tempb}{#3}}

\catcode\%=11
\def\html@spaces#1 #2{#1
\catcode\%=14
\makeatother

\newcommand{\todo}[1]{\textcolor{magenta}{[#1]}}
\newcommand{\jantodo}[1]{\textcolor{green}{Jan: #1}}
\newcommand{\alextodo}[1]{\textcolor{green}{Alex: #1}}
\newcommand{\stefanotodo}[1]{\textcolor{green}{Stefano: #1}}
\newcommand{\fdetodo}[1]{\textcolor{green}{FDE: #1}}

\newcommand{\citationneeded}{\textcolor{ForestGreen}{$^{\rm citation\;needed}$}}
\let\oldtextsigma\textsigma
\renewcommand{\textsigma}{\oldtextsigma\xspace}
\let\oldAA\AA
\renewcommand{\AA}{\text{\oldAA}\xspace}
\let\oldtextdegree\textdegree
\renewcommand{\textdegree}{\oldtextdegree\xspace}

\newcommand{\kms}{\ensuremath{\mathrm{km\,s^{-1}}}\xspace}
\newcommand{\Msun}{\ensuremath{{\rm M}_\odot}\xspace}
\newcommand{\Zsun}{\ensuremath{{\rm Z}_\odot}\xspace}
\newcommand{\yr}{\ensuremath{{\rm yr}}\xspace}
\newcommand{\Myr}{\ensuremath{{\rm Myr}}\xspace}
\newcommand{\Gyr}{\ensuremath{{\rm Gyr}}\xspace}
\newcommand{\peryr}{\ensuremath{{\rm yr^{-1}}}\xspace}
\newcommand{\Lsun}{\hbox{\,${\rm L}_\odot$}}
\newcommand{\mum}{\text{\textmu m}\xspace}
\newcommand{\kpc}{\text{kpc}\xspace}
\newcommand{\ZH}{\text{[Z/H]}\xspace}

\newcommandx{\lambdar}[2][1=R,2=]{\ensuremath{\lambda_{\rm {#1}}{#2}}\xspace}
\newcommand{\eps}{\ensuremath{\epsilon}\xspace}
\newcommand{\mstar}{\ensuremath{M_\star}\xspace}
\newcommand{\mdyn}{\ensuremath{M_\mathrm{dyn}}\xspace}
\newcommand{\re}{\ensuremath{R_\mathrm{e}}\xspace}
\newcommand{\vstar}{\ensuremath{v_\star}\xspace}
\newcommand{\vnai}{\ensuremath{v_{\NaI}}\xspace}
\newcommand{\sigmastar}{\ensuremath{\sigma_\star}\xspace}
\newcommand{\sigmaestar}{\ensuremath{\sigma_{\star,\mathrm{e}}}\xspace}
\newcommand{\vperc}[1]{\ensuremath{v_{#1}}\xspace}

\newcommand{\nelec}{\ensuremath{n_\mathrm{e}}\xspace}

\newcommandx{\fluxdcgs}[1][1=-20]{$\times 10^{[#1]}$~erg~s$^{-1}$~cm$^{-2}$~\AA$^{-1}$\xspace}
\newcommandx{\fluxcgs}[2][1=-20,2=\ensuremath{\times}]{${#2}10^{#1}$~erg~s$^{-1}$~cm$^{-2}$\xspace}
\newcommandx{\powercgs}[1][1=44]{$\times 10^{#1}$~erg~s$^{-1}$\xspace}
\newcommand{\Av}{\ensuremath{A_V}\xspace}

\newcommand{\jwst}{\textit{JWST}\xspace}
\newcommand{\hst}{\textit{HST}\xspace}

\newcommand{\ppxf}{{\sc ppxf}\xspace}
\newcommand{\fsps}{{\sc fsps}\xspace}
\newcommand{\prospector}{{\sc prospector}\xspace}
\newcommand{\bagpipes}{{\sc bagpipes}\xspace}
\newcommand{\beagle}{{\sc beagle}\xspace}
\newcommand{\emcee}{{\sc emcee}\xspace}
\newcommand{\cloudy}{{\sc cloudy}\xspace}
\newcommand{\fitsmap}{{\sc fitsmap}\xspace}
\newcommand{\eazy}{{\sc eazy}\xspace}
\newcommand{\empt}{{\sc eMPT}\xspace}
\newcommandx{\mappings}[1][1=]{{\sc mappings{#1}}\xspace}

\newcommand{\mediumjwstgn}{medium/\jwst~GN\xspace}
\newcommand{\mediumhstgn}{medium/\hst~GN\xspace}
\newcommand{\mediumhstgs}{medium/\hst~GS\xspace}
\newcommand{\mediumjwstgs}{medium/\jwst~GS\xspace}
\newcommand{\mediumjwstgsb}{medium/\jwst~GS~1180\xspace}
\newcommand{\ultradeep}{ultradeep~GS\xspace}

\newcommand{\Lyalpha}{\text{Ly\textalpha}\xspace}
\newcommand{\Halpha}{\text{H\textalpha}\xspace}
\newcommand{\Hbeta}{\text{H\textbeta}\xspace}
\newcommand{\Hgamma}{\text{H\textgamma}\xspace}
\newcommand{\Hdelta}{\text{H\textdelta}\xspace}
\newcommand{\Hepsilon}{\text{H\textepsilon}\xspace}
\newcommand{\Hzeta}{\text{H\textdzeta}\xspace}
\newcommand{\Paalpha}{\text{Pa\textalpha}\xspace}
\newcommand{\Pabeta}{\text{Pa\textbeta}\xspace}
\newcommand{\Pagamma}{\text{Pa\textgamma}\xspace}
\newcommand{\Padelta}{\text{Pa\textdelta}\xspace}

\newcommandx{\permittedEL}[6][1=O,2=III,3=,4=,5=,6=]{\text{{#1}\,{\sc {#2}}{#3}{#4}{#5}{#6}}\xspace}
\newcommandx{\semiforbiddenEL}[6][1=O,2=III,3=,4=,5=,6=]{\text{{#1}\,{\sc{#2}}]{#3}{#4}{#5}{#6}}\xspace}
\newcommandx{\forbiddenEL}[6][1=O,2=III,3=,4=,5=,6=]{\text{[{#1}\,{\sc{#2}}]{#3}{#4}{#5}{#6}}\xspace}

\newcommand{\EW}[1]{\text{EW(#1)}\xspace}

\newcommand{\HI}{\permittedEL[H][i]}
\newcommand{\HII}{\permittedEL[H][ii]}

\newcommand{\NV}{\permittedEL[N][v]}
\newcommandx{\NVL}[1][1=1243]{\permittedEL[N][v][\textlambda][#1]}
\newcommandx{\NVall}{\permittedEL[N][v][\textlambda][\textlambda][1239,][1243]}

\newcommandx{\CIIL}[1][1=232x]{\semiforbiddenEL[C][ii][\textlambda][#1]}
\newcommandx{\CIIall}{\semiforbiddenEL[C][ii][\textlambda][\textlambda][2323.5--][2328.1]}

\newcommand{\NIV}{\semiforbiddenEL[N][iv]}
\newcommandx{\NIVL}[1][1=1486]{\semiforbiddenEL[N][iv][\textlambda][#1]}

\newcommand{\CIV}{\permittedEL[C][iv]}
\newcommandx{\CIVL}[1][1=1550]{\permittedEL[C][iv][\textlambda][#1]}
\newcommand{\CIVall}{\permittedEL[C][iv][\textlambda][\textlambda][1549,][1551]}

\newcommand{\HeII}{\permittedEL[He][ii]}
\newcommandx{\HeIIL}[1][1=1640]{\permittedEL[He][ii][\textlambda][#1]}

\newcommand{\semiOIII}{\semiforbiddenEL[O][iii]}
\newcommandx{\semiOIIIL}[1][1=1666]{\semiforbiddenEL[O][iii][\textlambda][#1]}
\newcommand{\semiOIIIall}{\semiforbiddenEL[O][iii][\textlambda][\textlambda][1661,][1666]}

\newcommand{\NIII}{\semiforbiddenEL[N][iii]}
\newcommandx{\NIIIL}[1][1=1750]{\semiforbiddenEL[N][iii][\textlambda][#1]}
\newcommand{\NIIIall}{\semiforbiddenEL[N][iii][\textlambda][\textlambda][1747--][1754]}

\newcommandx{\CIII}{\semiforbiddenEL[C][iii]}
\newcommandx{\CIIIL}[1][1=1909]{\semiforbiddenEL[C][iii][\textlambda][#1]}
\newcommand{\CIIIall}{\semiforbiddenEL[C][iii][\textlambda][\textlambda][1907,][1909]}

\newcommand{\NeIV}{\forbiddenEL[Ne][iv]}
\newcommandx{\NeIVL}[1][1=2424]{\forbiddenEL[Ne][iv][\textlambda][#1]}
\newcommand{\NeIVall}{\forbiddenEL[Ne][iv][\textlambda][\textlambda][2422,][2424]}

\newcommand{\MgII}{\permittedEL[Mg][ii]}
\newcommandx{\MgIIL}[1][1=2803]{\permittedEL[Mg][ii][\textlambda][#1]}
\newcommand{\MgIIall}{\permittedEL[Mg][ii][\textlambda][\textlambda][2796,][2803]}

\newcommand{\NeV}{\forbiddenEL[Ne][v]}
\newcommandx{\NeVL}[1][1=3426]{\forbiddenEL[Ne][v][\textlambda][#1]}
\newcommand{\NeVall}{\forbiddenEL[Ne][v][\textlambda][\textlambda][3346,][3426]}

\newcommand{\OII}{\forbiddenEL[O][ii]}
\newcommandx{\OIIL}[1][1=3727]{\forbiddenEL[O][ii][\textlambda][#1]}
\newcommand{\OIIall}{\forbiddenEL[O][ii][\textlambda][\textlambda][3726,][3729]}

\newcommand{\NeIII}{\forbiddenEL[Ne][iii]}
\newcommandx{\NeIIIL}[1][1=3869]{\forbiddenEL[Ne][iii][\textlambda][#1]}
\newcommand{\NeIIIall}{\forbiddenEL[Ne][iii][\textlambda][\textlambda][3869,][3968]}

\newcommand{\OIII}{\forbiddenEL[O][iii]}
\newcommandx{\OIIIL}[1][1=5007]{\forbiddenEL[O][iii][\textlambda][#1]}
\newcommand{\OIIIall}{\forbiddenEL[O][iii][\textlambda][\textlambda][4959,][5007]}

\newcommand{\NaI}{\permitted[Na][i]}
\newcommandx{\NaIL}[1][1=5890]{\permittedEL[Na][i][\textlambda][#1]}
\newcommand{\NaIall}{\permittedEL[Na][i][\textlambda][\textlambda][5890,][5896]}

\newcommand{\OI}{\forbiddenEL[O][i]}
\newcommandx{\OIL}[1][1=6300]{\forbiddenEL[O][i][\textlambda][#1]}
\newcommand{\OIall}{\forbiddenEL[O][i][\textlambda][\textlambda][6300,][6363]}

\newcommand{\HeI}{\permittedEL[He][i]}
\newcommandx{\HeIL}[1][1=10830]{\permittedEL[He][i][\textlambda][#1]}

\newcommand{\NII}{\forbiddenEL[N][ii]}
\newcommandx{\NIIL}[1][1=6584]{\forbiddenEL[N][ii][\textlambda][#1]}
\newcommand{\NIIall}{\forbiddenEL[N][ii][\textlambda][\textlambda][6548,][6583]}

\newcommand{\SII}{\forbiddenEL[S][ii]}
\newcommand{\SIIL}[1][1=6716]{\forbiddenEL[S][ii][\textlambda][#1]}
\newcommand{\SIIall}{\forbiddenEL[S][ii][\textlambda][\textlambda][6716,][6731]}

\newcommandx{\OIIAuL}[1][1=7325]{\forbiddenEL[O][ii][\textlambda][#1]}
\newcommand{\OIIAuall}{\forbiddenEL[O][ii][\textlambda][\textlambda][7319--][7332]}

\newcommand{\SIII}{\forbiddenEL[S][iii]}
\newcommandx{\SIIIL}[1][1=9532]{\forbiddenEL[S][iii][\textlambda][#1]}
\newcommand{\SIIIall}{\forbiddenEL[S][iii][\textlambda][\textlambda][9069,][9532]}

\newcommandx{\SIIAuL}[1][1=10290]{\forbiddenEL[S][ii][\textlambda][#1]}
\newcommand{\SIIAuall}{\forbiddenEL[S][ii][\textlambda][\textlambda][10290--][10373]}

\newcommand{\hda}{\ensuremath{\mathrm{H\text{\textdelta}_A}}\xspace}
\newcommand{\hga}{\ensuremath{\mathrm{H\text{\textgamma}_A}}\xspace}

\defcitealias{bunker+2023b}{B24}
\defcitealias{JADES_DR3}{DE25}
\defcitealias{DR4_paper1}{Paper I}

\title[JADES DR4 Paper II]{JADES Data Release 4 - Paper \textsc{II}: Data reduction, analysis and emission-line fluxes of the complete spectroscopic sample}

\author[Jan Scholtz, Stefano Carniani]{\parbox[h]{\textwidth}{
J. Scholtz$^{1,2}$\thanks{E-mail: js2685@cam.ac.uk}\thanks{These authors contributed to this work equally.},
S. Carniani$\dagger$\thanks{E-mail: stefano.carniani@sns.it}$^{3}$,
E. Parlanti$^{3}$,
F. D'Eugenio$^{1,2}$,
E. Curtis-Lake$^{4}$,
P. Jakobsen$^{5,6}$,
A. J. Bunker$^{7}$,
A. J. Cameron$^{7}$,
S. Arribas$^{8}$, 
W. M. Baker$^{9}$,
S. Charlot$^{10}$ , 
J. Chevellard$^{7}$,
C. Circosta$^{11}$,
M. Curti$^{12}$,  
Q. Duan$^{1,2}$,
D. J.\ Eisenstein$^{13}$,
K. Hainline$^{14}$,
Z. Ji$^{14}$,
B. D\ Johnson$^{13}$, 
G. C. Jones$^{1,2}$,
N. Kumari$^{15}$,
R. Maiolino$^{1,2,16}$,
M. V. Maseda$^{17}$,
M. Perna$^{8}$, 
P. G. Pérez-González$^{8}$, 
T. Rawle$^{18}$,
M. Rieke$^{14}$,
P. Rinaldi$^{19}$,
B. Robertson$^{20}$,
A. Saxena$^{7,16}$,
I. Shivaei$^{8}$,
M. S. Silcock$^{4}$,
Y. Sun$^{14}$,
B. Rodr\'iguez Del Pino$^{8}$,
S. Tacchella$^{1,2}$,
H. \"Ubler$^{21}$,
G. Venturi$^{3}$,
C. C. Williams$^{22}$,
C. N. A. Willmer$^{14}$,
C. Willott$^{23}$,
J. Witstok$^{5,6}$
}\vspace{0.4cm}
\\
$^{1}$Kavli Institute for Cosmology, University of Cambridge, Madingley Road, Cambridge, 
CB3 0HA, UK\\
$^{2}$Cavendish Laboratory, University of Cambridge, 19 JJ Thomson Avenue, Cambridge CB3 0HE, UK\\
$^{3}$ Scuola Normale Superiore, Piazza dei Cavalieri 7, I-56126 Pisa, Italy\\
$^{4}$Centre for Astrophysics Research, Department of Physics, Astronomy and Mathematics, University of Hertfordshire, Hatfield AL10 9AB, UK \\
$^{5}$ Cosmic Dawn Center (DAWN), Copenhagen, Denmark\\
$^{6}$ Niels Bohr Institute, University of Copenhagen, Jagtvej 128, DK-2200, Copenhagen, D \\
$^{7}$ University of Oxford, Department of Physics, Denys Wilkinson Building, Keble Road, Oxford OX13RH, United Kingdom\\
$^{8}$ Centro de Astrobiolog\'ia (CAB), CSIC–INTA, Cra. de Ajalvir Km.~4, 28850- Torrej\'on de Ardoz, Madrid, Spain\\
$^{9}$ DARK, Niels Bohr Institute, University of Copenhagen, Jagtvej 155A, DK-2200 Copenhagen, Denmark\\
$^{10}$ Sorbonne Universit\'e, CNRS, UMR 7095, Institut d'Astrophysique de Paris, 98 bis bd Arago, 75014 Paris, France\\
$^{11}$ Institut de Radioastronomie Millimétrique (IRAM), 300 Rue de la Piscine, 38400 Saint-Martin-d'Hères, France\\
$^{12}$ European Southern Observatory, Karl-Schwarzschild-Strasse 2, 85748 Garching, Germany \\
$^{13}$ Center for Astrophysics $|$ Harvard \& Smithsonian, 60 Garden St., Cambridge MA 02138 USA\\
$^{14}$ Steward Observatory, University of Arizona, 933 N. Cherry Avenue, Tucson, AZ 85721, USA \\
$^{15}$ AURA for European Space Agency, Space Telescope Science Institute, 3700 San Martin Drive. Baltimore, MD, 21210 \\
$^{16}$ Department of Physics and Astronomy, University College London, Gower Street, London WC1E 6BT, UK\\
$^{17}$  Department of Astronomy, University of Wisconsin-Madison, 475 N. Charter St., Madison, WI 53706 USA\\
$^{18}$ European Space Agency (ESA), European Space Astronomy Centre (ESAC), Camino Bajo del Castillo s/n, 28692 Villafranca del Castillo, Madrid, Spain \\
$^{19}$  Space Telescope Science Institute, 3700 San Martin Drive, Baltimore, Maryland 21218, USA\\
$^{20}$  Department of Astronomy and Astrophysics, University of California, Santa Cruz, 1156 High Street, Santa Cruz, CA 95064, USA \\
$^{21}$  Max-Planck-Institut f\"ur extraterrestrische Physik (MPE), Gie{\ss}enbachstra{\ss}e 1, 85748 Garching, Germany \\
$^{22}$  NSF National Optical-Infrared Astronomy Research Laboratory, 950 North Cherry Avenue, Tucson, AZ 85719, USA\\
$^{23}$  NRC Herzberg, 5071 West Saanich Rd, Victoria, BC V9E 2E7, Canada \\
\vspace{-1 cm}
}

\date{Accepted XXX. Received YYY; in original form ZZZ}

\pubyear{2025}

\begin{document}
\label{firstpage}
\pagerange{\pageref{firstpage}--\pageref{lastpage}}
\maketitle

\begin{abstract}
We present the fourth data release of JADES, the \jwst Advanced Deep Extragalactic Survey, providing deep spectroscopic observations in the two GOODS fields. A companion paper presents the target selection, spectroscopic redshifts and success rates, and in this paper, we discuss the data reduction and present emission line flux measurements. The spectroscopy in this work consists of medium-depth, deep and ultradeep NIRSpec/MSA spectra of 5,190 targets, covering the spectral range $0.6\text{--}5.5$~\mum and observed with both the low-dispersion prism ($R=30\text{--}300$) and all three medium-resolution gratings ($R=500\text{--}1,500$). We describe the data reduction, analysis and description of the data products included in this data release. In total, we measured 3,297 robust redshifts out of  5,190 targets, spanning a redshift range from $z=0.5$ up to $z=14.2$, including 974 at $z>4$. This data release includes 1-d and 2-d fully reduced spectra with 3 and 5 pixel extractions, with slit-loss corrections and background subtraction optimized for point sources. Furthermore, we provide redshifts and $S/N>5$ emission-line flux catalogues for the prism and grating spectra, as well as new guidelines to use these data products. Lastly, we are launching a new JADES Online Database, designed to enable quick selection and browsing of this data release. Altogether, these data provide the largest statistical sample to date to characterise the properties of galaxy populations across Cosmic time. 

\end{abstract}




\section{Introduction}\label{s.intro}

Since the launch of \jwst nearly four years ago, the observatory has revolutionised our view of the Universe and early galaxy evolution. Many of these discoveries were possible thanks to the amazing capabilities and sensitivity of the NIRSpec instrument \citep{jakobsen+2022}. These discoveries include e.g.: confirmation of $z>10$ galaxies \citep{curtis-lake+2023,arrabal-haro+2023,wang+2023, carniani+2024,Castellano+2024, Harikane+2024, Hsiao+2024, Kokorev+2025, Naidu+2025, witstok+2025}, many with detected emission lines 
\citep[][]{bunker+2023a, Castellano+2024, deugenio_gsz12, Naidu+2025, Scholtz+2025}, from which chemical abundances have been derived \citep[e.g.,][]{curti+2023a,nakajima+2023, isobe+2025, Ji+2025_NO}, the discovery of existance of massive, quiescent and old galaxies at $z=3\text{--}5$ \citep[e.g.,][]{carnall+2023b,glazebrook+2023,2024arXiv240503744P, Baker+2025, deGraaf2025_nature, Weibel2025}; the first `mini-quenched' galaxies \citep{looser+2024,strait+2023, Baker+2025b}; neutral-phase outflows in massive quiescent galaxies \citep{belli+2023,deugenio+2023c,davies+2024, Valentino+2025}; the discovery of metal-poor active galactic nuclei \citep[e.g.,][]{kocevski+2023,uebler+2023, maiolino+2023b, juodzbalis+2025, scholtz+2025b, maiolino2025qso1};
the most distant active galactic nuclei \citep[AGN;][]{maiolino+2023a,goulding+2023};
and even tentative evidence of the first generation of stars \citep[][]{vanzella+2023, maiolino+2023b, Nakajima2025,Vanzella2025}. However, for spectroscopy, the large samples are either still relatively shallow (such as Rubies, WIDE; \citealt{degraaff+2025, maseda+2024}), or the deep observations have small number statistics, limited to tens to hundred of objects \citep{looser+2023,curti+2023b,nakajima+2023}. The key to large scale characterization of the galaxy population at high redshift is a large sample of deep spectroscopic observations, which are able to disentangle the complex mechanisms in galaxy evolution. Indeed, at low redshift ($z<1$), there has been a significant investment of resources to create samples of hundreds of thousands of spectroscopic observations of galaxies \citep[e.g.,][]{abazajian+2009,driver+2018,desi+2016a}. Studies focusing on multiple physical properties of galaxies at once \citep[e.g.,][]{kauffmann+2003a,kauffmann+2003b,peng+2010,graves+faber2010}, or deploying new machine-learning algorithms  \citep[e.g.,][]{bluck+2022,baker+2022,barsanti+2023,walmsley+2023,koller2025}, have made enormous progress in galaxy evolution, unveiling the links between star-formation, stellar ages, supermassive black-hole mass, metallicity, morphology and environment. Furthermore, within the last two decades, there has been a push for large spectroscopic ground-based surveys which have enabled the community to study several hundreds of galaxies up to z=4 \citep[e.g.,][]{wisnioski+2015,stott+2016,kriek+2015}. Within the next few years, new facilities and surveys will expand this number to up to 1 million galaxies at $z \lesssim 4$ \citep{dalton+2012,tamura+2016,dejong+2019,maiolino+2020}. However, currently only \jwst can obtain deep rest-frame optical and UV spectra for a large sample of galaxies above z$>$4, which is essential to study the early phases of galaxy evolution.

The major progress at z$>$4 has been enabled by the NIRSpec Micro Shutter Assembly \citep[MSA,][]{ferruit+2022}, which was designed to observe more than two hundred targets with a single pointing. This high multiplicity slit spectroscopy is combined with the unprecedented collecting area of the \jwst mirror, low background and nominal wavelength coverage up to 5.5 $\mu$m. As a result, we are now able to observe rest-frame optical and UV emission of a large sample of galaxies $z>4$  \citetext{e.g., \citealp{treu+2022}, \citealp{bezanson+2022}, \citealp{fujimoto+2023}, \citealp{oesch+2023}, \citealp{bunker+2023b}, hereafter: \citetalias{bunker+2023b}}.

With these goals and capabilities in mind,  the \jwst Advanced Deep Extragalactic Survey \citep[JADES;][]{eisenstein+2023a}, took spectra of more than 2207 galaxies at redshifts $z>3$, enabling us to shift our focus from discovering high redshift galaxies to statistically investigating their properties. To maximise the efficiency of this process, JADES is a combined collaboration of \jwst NIRCam and NIRSpec GTO teams to fully exploit the combination of imaging and spectroscopic data. The JADES survey strategy is split into the three tiers: medium-depth, deep-depth and ultradeep-depth \citep[for the shallowest and widest tier of the NIRSpec GTO see][]{maseda+2024}, targeting the two GOODS fields \citep[Great Observatories Origins Deep Survey;][]{giavalisco_goods_2004}. While the medium tiers were observed in both GOODS South and North fields (hereafter, GOODS-S and GOODS-N), the two deepest tiers focused on GOODS-S only. There were three previous data releases (DRs): i) DR 1, which is split between imaging \citep{rieke_jades_2023} and spectroscopy \citepalias{bunker+2023b}; ii) DR2, which provided new GOODS-S imaging \citep{eisenstein+2023b}; and iii) DR3, which represented a significant proportion of the total survey imaging and spectroscopic data across GOODS-S and GOODS-N \citep[][referred to as \citetalias{JADES_DR3}]{JADES_DR3}. 

In this work and along with the Paper I \citep[][ from now on \citetalias{DR4_paper1}]{DR4_paper1}, we present the final data release of the spectroscopic part of JADES with final medium and deep observations in GOODS-S, as well as full re-reduction and re-analysis of the previous observations presented in \citetalias{JADES_DR3}. In this data release, we provide fully reduced and calibrated 1-d and 2-d spectra, NIRCam image cutouts for our spectroscopic targets,  as well as measurements of redshift and emission-line fluxes across two distinct extractions \footnote{Available on the JADES website \url{https://jades-survey.github.io/scientists/data.html}.}. The sample selection, redshifts and recovered redshift success rates of the JADES spectroscopic survey are described in \citetalias{DR4_paper1}. We briefly describe the observing strategy in \S~\ref{sec:observations} and our new data reduction in \S~\ref{s.datared}. We then outline the measurements of spectroscopic redshifts and line fluxes (\S~\ref{s.r100}--\ref{s.r1000}).
In Sections~\ref{s.quality} and~\ref{s.limitations} we present an assessment of the data products and guidelines for their use. We conclude with a short summary and brief outlook (Section~\ref{s.conclusions}). 
Throughout this work, we use the AB~magnitude system \citep{oke+gunn1983}.

\section{NIRSpec/MSA observations}
\label{sec:observations}

\begin{table*}
\tiny
\begin{center}
\caption{Summary of JADES NIRSpec/MSA observations. Under each disperser, we report the (minimum/mean/maximum) exposure times; the minimum exposure time can be 0, due to disobedient shutters (for PRISM) and for protecting high-priority targets from overlap (for the gratings).}
\label{tab:obssummary}
\begin{tabular}{lccclcccccc}
  \hline
PID          &  Field  & Depth  & Selection & Tier name                     & PRISM & G140M & G235M & G395M & Targets & Initial Release \\ 
             &         &        &           &                               &   [h] &   [h] &   [h] &   [h] &         &         \\ 
\hline
1210     & GOODS-S &   Deep & HST/JWST &        \verb|goods-s-deephst| & ( 9.2/16.5/27.7) & ( 2.3/ 4.1/ 7.0) & ( 2.3/ 4.1/ 7.0) & ( 2.3/ 4.1/ 7.0) & 253 & \citetalias{bunker+2023b} \\
\hline
1180 & GOODS-S & Medium &      HST &      \verb|goods-s-mediumhst| & ( 0.9/ 1.0/ 4.3) & ( 0.9/ 1.0/ 4.3) & ( 0.9/ 1.0/ 4.3) & ( 0.9/ 1.0/ 4.3) & 677 & \citetalias{JADES_DR3} \\
1180     & GOODS-S & Medium &     JWST & \verb|goods-s-mediumjwst1180| & ( 0.3/ 2.1/ 5.2) & ( 0.9/ 1.8/ 4.3) & ( 0.9/ 1.8/ 4.3) & ( 0.9/ 1.8/ 4.3) & 533 & " \\
1181     & GOODS-N & Medium &      HST &      \verb|goods-n-mediumhst| & ( 0.6/ 2.0/ 6.9) & ( 0.9/ 1.0/ 3.5) & ( 0.9/ 1.0/ 3.5) & ( 0.9/ 1.0/ 3.5) &  853 & " \\
1181     & GOODS-N & Medium &     JWST &     \verb|goods-n-mediumjwst| & ( 0.3/ 1.6/ 5.2) & ( 0.9/ 1.7/ 5.2) & ( 0.9/ 1.7/ 5.2) & ( 0.9/ 1.7/ 5.2) & 950 & " \\
3215     & GOODS-S &   Ultradeep &      JWST &      \verb|goods-s-ultradeep| & ( 2.8/32.4/61.6) & ( 2.8/ 7.7/11.2) &        ---       & (11.2/23.0/33.6) & 228 & " \\
\hline
1286     & GOODS-S & Medium &     JWST &     \verb|goods-s-mediumjwst| & ( 0.5/ 2.1/ 2.2) & ( 0.7/ 2.1/ 2.2) & ( 0.9/ 2.4/ 2.6) & ( 0.9/ 2.4/ 2.6) & 1490 & This work \\
1287     & GOODS-S & Deep   &     JWST &      \verb|goods-s-deepjwst|         &   ( 9.2/16.5/27.7) & ( 2.3/ 4.1/ 7.0) & ( 2.3/ 4.1/ 7.0) & ( 2.3/ 4.1/ 7.0) & 235 &      " \\
  \hline
\end{tabular}
\end{center}
\end{table*}

All observations used NIRSpec in Multi-Object Spectroscopy mode, with the NIRSpec/MSA \citep{ferruit+2022}. The MSA configurations were planned using the prioritisation schemes detailed in \citetalias{DR4_paper1}, and the strategy described in \cite{eisenstein+2023a}. The sample selection of our GTO programme is designed to understand the star-formation history at high-redshifts ($z\gtrsim5.7)$, where galaxies are predominantly selected based on their rest-frame ultraviolet properties, and the mass-assembly history at lower redshifts where objects are selected from their rest-frame optical.  Rare or interesting targets are also added as `oddballs'.  Objects with lower number densities are generally given a higher priority, while more numerous galaxies occupy the lower priority classes.

Here we briefly describe the observing strategy. Each visit is planned with a set of target acquisition (TA) objects (stars and compact galaxies), which were identified in the same images as those used to find positions of the science targets, to ensure the astrometry of the target acquisition sources and the science observations is identical. The TA targets were visually inspected to ensure they met the sufficient quality: compact, symmetric, and did not have colour gradients or nearby sources. All TAs used the NIRSpec CLEAR filter and the longest readout time (mode NRSRAPIDD6) since the TA objects in the GOODS fields are faint. Further details on the TA setups are provided in \cite{eisenstein+2023a}. All TAs performed for our observations have been successful.

As described in detail by \citetalias{JADES_DR3}, there were several technical issues with JADES NIRSpec observations, which resulted in some visits being skipped or partially collected either due to guide-star acquisition or re-acquisition failures from the Fine Guidance Sensor and `shorts' \citep[electrical short circuits with the NIRSpec MSA;][]{rawle+2022}. For partially collected observations of MSA configurations, the strategy was to return to the MSA configuration one year later to ensure successful completion of our programme with the selected targets. For cases where no data were obtained for an MSA configuration, we have re-planned at either the same or at a different orientation. 

In Table \ref{tab:obssummary}, we present the complete summary of JADES NIRSpec observations, including the typical integration time (we provide the minimum, mean and maximum integration time for each spectral configuration). As our observing strategy, we use the \texttt{nrsirs2} readout mode \citep[improved reference sampling and subtraction, or \texttt{irs$^2$};][]{rauscher+2012,rauscher+2017}. We generally describe the observations with a label structured as `depth/selection', where depth is either `Medium', `Deep', or `Ultradeep', and selection is either `HST' or `JWST', depending on how the majority of targets was selected; these labels are then `translated' into the \verb|TIER| column in the published tables, and are part of the file names for the published spectra.

The full details of what is new to this data release are provided in Section 2.2 of \citetalias{DR4_paper1}, though we provide here an overview of the completely new sets of observations being included.  In the following subsections, we provide additional details on the new observations being published in this data release; the rest of the descriptions are in \citetalias{JADES_DR3} and \citetalias{bunker+2023b}. The DR4 electronic files provide more details, such as observation date and actual integration times per target.

We are including in this data release two tiers that were previously unpublished: 

\begin{enumerate}
    \item \textbf{1286: GOODS-S Medium/JWST}: Program ID 1286 is the main GOODS-S Medium/JWST program. These observations were planned in the same way as those of GOODS-N Medium/JWST, each with three sub-pointings separated by $\approx 1$~arcsec. Compared to \citetalias{JADES_DR3}, all observations from this programme are now included, with the new observations from October and December 2023. For this visit, the integration times per sub-pointing in some dispersers were reduced from 3.1 to 2.7 ksec to fit within the available time allocation.

    \item \textbf{1287: GOODS-S Deep/JWST}: Program ID 1287 is the main GOODS-S Deep/JWST programme, which shares the observational setup with GOODS-S Deep/HST, but different pointings with three dither positions, each achieving 9.3h on source exposure, as well as 2.3h per grating per dither position. Given the design of the observations, many targets were observed up to 27 h in Prism and 7 h in the gratings.
\end{enumerate}


\medskip


With this release, we include a table containing the redshifts, target information used for mask preparation, and emission line measurements.  We label this table as the \texttt{Master}.
We summarise the main properties of each target in the \texttt{Master} table. The columns of this table are described in Table \ref{tab:Master}. We note that the order of the targets in the \texttt{Obs\_info} table is the same as subsequent tables for PRISM and R1000 flux measurements described in section \ref{s.products}.

\begin{table}\label{tab:Master}
\begin{center}
\caption{Structure of the \texttt{Obs\_info} target table for the Data Release 4.}
\begin{tabular}{ll}
    \hline
    Column name               & Description \\
    \hline
    \texttt{Unique\_ID}       & Unique ID of the source in the survey  \\
    \texttt{PID}              & JWST Programme ID \\
    \texttt{TIER}             & Name of subset \\
    \texttt{TIER\_old}         & Old name of subset in previous DR \\
    \texttt{NIRSpec\_ID}      & ID of the target in eMPT \\
    \texttt{NIRCam\_DR5\_ID}  & NIRCam ID from upcoming DR5  \\
    \texttt{NIRCam\_DR3\_ID}  & NIRCam ID from DR3 (\citetalias{JADES_DR3})\\
    \texttt{ObsDate}          & Date of observations \\
    \texttt{RA\_TARG}         & Target right ascension [degrees] \\
    \texttt{Dec\_TARG}        & Target declination [degrees] \\
    \texttt{x\_offset}        & Intra-shutter target offset [arcsec] \\
    \texttt{y\_offset}        & Intra-shutter target offset [arcsec] \\
    \texttt{Field}            & Name of field (GS or GN) \\
    \texttt{GSa}              & Selection method used for GOODS-S field - a$^{*}$ \\
    \texttt{GSb}              & Selection method used for GOODS-S field - b$^{*}$ \\
    \texttt{Priority}         & Target Priority \\
    \texttt{PC\_empt}         & Priority class within empt software\\
    \texttt{Priority}         & Priority class in target assignment process\\
    \texttt{oddball\_source}  & Source catalogue used for oddball selection\\
    \texttt{PC\_eMPT\_pre\_oddball} & Priority class within empt software before \\
    \dots                     & it was selected as oddball \\
    \texttt{priority\_pre\_oddball} & Priority class in target assignment process \\
    \dots                     & before it was selected as oddball \\
    \texttt{probability\_allocated} & Probability of being allocated an MSA shutter\\
    \texttt{F444W\_gold\_DR3} & Gold sample based on \texttt{F444W} in DR3 \\
    \dots                     & photometry $^{*}$\\
    \texttt{UV\_gold\_DR3}    & Gold sample based on UV magnitude$^{*}$ \\
    \dots                     &  in DR3 photometry$^{*}$\\
    \texttt{F444W\_gold\_DR5\_beta} & Gold sample based on \texttt{F444W} in upcoming \\
    \dots                     & DR5 photometry$^{*}$\\
    \texttt{UV\_gold\_DR5\_beta}    & Gold sample based on UV magnitude \\
    \dots                     &  in upcoming DR5 photometry$^{*}$\\
    \texttt{assigned\_Prism}  & \texttt{True} if has prism observations \\
    \dots                     & \dots \\
    \texttt{assigned\_G395H}  & \texttt{True} if has G395H observations \\
    \texttt{nDither\_Pr}      & Number of dithers for prism \\
    \texttt{nDither\_Gr}      & Number of dithers for gratings \\
    \texttt{nInt\_Prism}      & Number of integrations for prism \\
    \dots                     & \dots \\
    \texttt{nInt\_G395H}     & Number of integrations for G395H \\
    \texttt{tExp\_PRISM}      & Exposure time for prism [s] \\
    \dots                     & \dots \\
    \texttt{tExp\_G395H}      & Exposure time for F290LP/G395H [s] \\
    \texttt{z\_Spec}          & Redshift (both prism and gratings)      \\
    \texttt{z\_Spec\_flag}    & Redshift flag (both prism and gratings) \\
    \texttt{z\_paper}        & redshift from previsouly published work  \\
    \texttt{z\_paper\_name}   & Name of the object in the previous work  \\
    \texttt{z\_paper\_ref}    & Citation to the previous work  \\
    \texttt{z\_PRISM}         & Prism-based redshift                    \\
    \texttt{z\_R1000}         & Grating-based redshift                    \\
    \texttt{z\_R1000n}         & Number of emission lines used to identify \\
     & Grating-based redshift                    \\
    \texttt{1500A\_flux}         & 1500 \AA flux  \\
    \texttt{1500A\_flux\_err}         & 1500 \AA flux  error\\
    \texttt{MUV}         & Absolute UV magnitude\\
    \texttt{MUV\_err\_u}         & 1500 \AA flux upper error\\
    \texttt{MUV\_err\_l}         & 1500 \AA flux upper lower\\
    \hline
\end{tabular}
\par $^{*}$ for the selection method refer to \citetalias{DR4_paper1}. 
\end{center}
\end{table}

\section{NIRSpec/MSA Data Reduction}\label{s.datared}

\begin{figure*}
  \includegraphics[trim={0 20cm 0cm 0},clip, width=\textwidth]{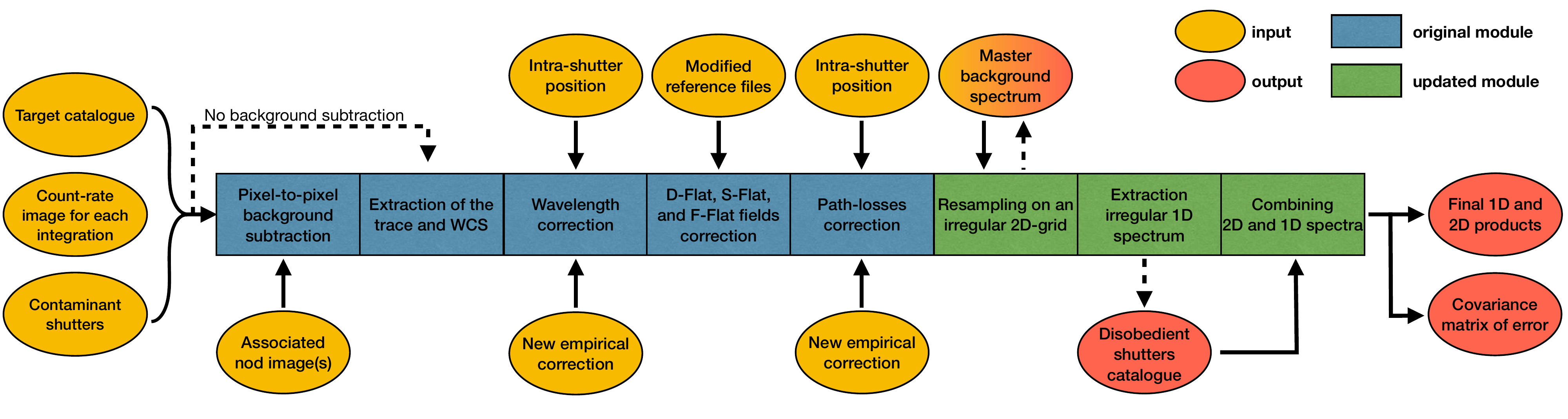}
  \caption{Outline of the overall workflow of the GTO NIRSpec/MSA pipeline. The blue and green indicate the original and updated modules in the workflow. The yellow and red regions show the input and output of the pipeline, respectively.}\label{f.pipeline}
\end{figure*}

The data processing for this release is based on the latest version of the NIRSpec GTO pipeline version 5.1, with the Calibration Reference Data
System context file of jwst1413.pmap. The overall structure of the pipeline is similar to that described in 
\citetalias{bunker+2023b} and \citetalias{JADES_DR3}, although several steps have been updated and automated to minimize user intervention and ensure consistent results across the various tiers. This section provides a detailed description of the main steps included in the data processing workflow (Figure~\ref{f.pipeline}).

At the core of the NIRSpec GTO pipeline is the extraction of calibrated spectra from NIRSpec count-rate images retrieved directly from the MAST archive. This process relies mainly on the NIRSpec Instrument Performance
Simulator pipeline software \citep[NIPS;][]{Dorner2016} developed by the ESA NIRSpec Science Operations Team.  NIPS was designed with three primary objectives: (a) to process pre-JWST simulations \citep{Giardino2019}, (b) to support data processing during commissioning \citep[e.g.,][]{Lutzgendorf2022}, and (c) to develop and validate algorithms for inclusion in the standard STScI pipeline. These algorithms are described in \citet{AlvesdeOliveira2018} and \citet{Ferruit2022}, and have all been incorporated into the STScI pipeline \citep{AlvesdeOliveira2018}. Due to licensing restrictions, limited resources for user support, documentation, and code maintenance, ESA-developed Python packages are not publicly released. 

Over the past three years, the NIRSpec GTO team has updated or replaced several components of the NIPS pipeline to optimize the quality of the final data products in alignment with the scientific objectives of the JADES program. In addition, the data processing workflow has been fine-tuned based on a default MSA configuration that has been used in all JADES tiers and programs. This configuration involves allocating three contiguous shutters perpendicular to the spectral dispersion direction for each target, followed by a 3-point nodding pattern (0, $\pm$1) along the resulting slitlet. All algorithms developed for the NIRSpec GTO pipeline are available upon request.

\begin{figure}
\centering
  \includegraphics[trim={1.5cm 1.3cm 0 1.3cm},clip, width=0.4\textwidth]{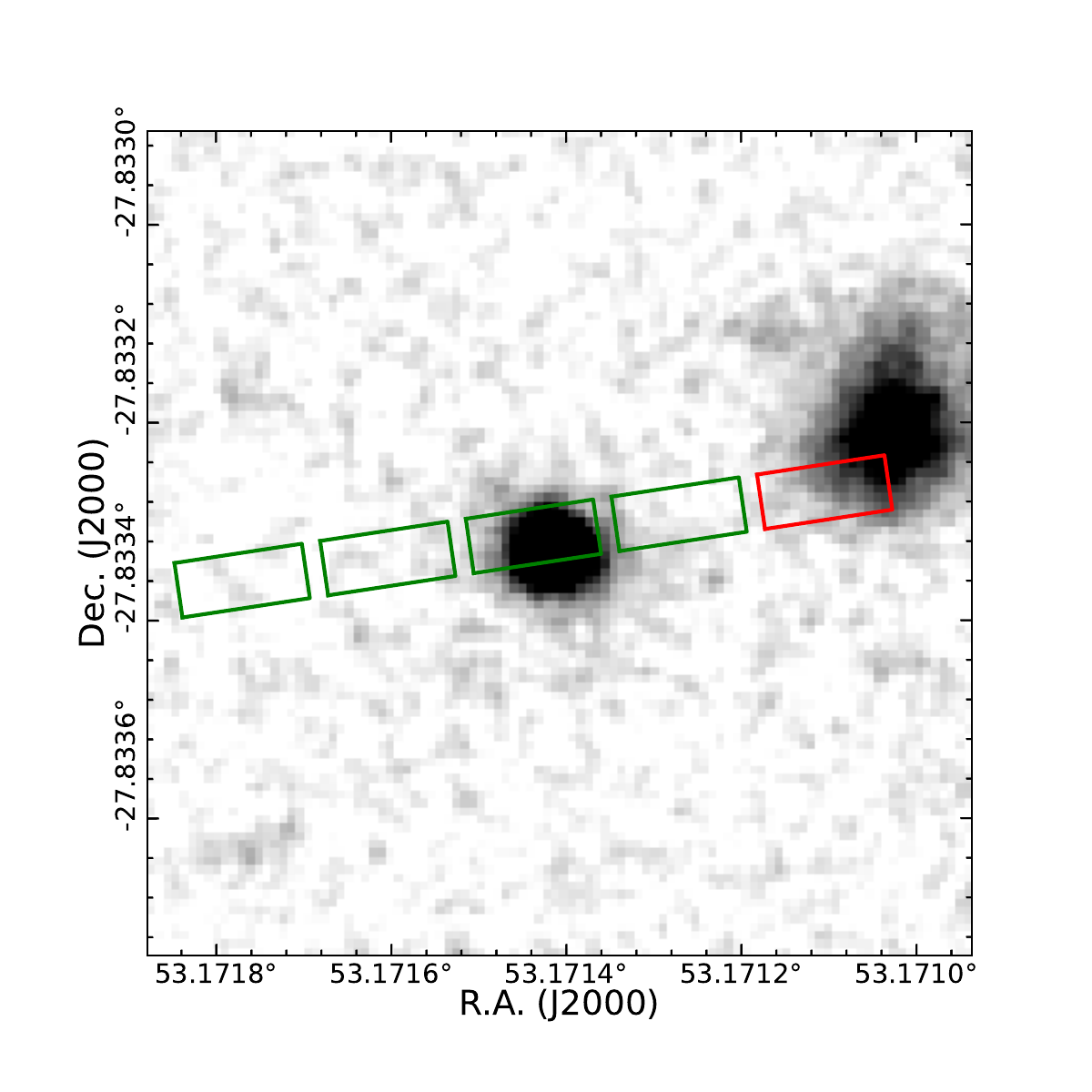}
  \caption{F277W NIRCam $5^{\prime\prime}\times5^{\prime\prime}$ cutout of the JADES target 86727 in the tier goods-s-mediumjwst.  
  The MSA shutters corresponding to the 3-point nod pattern are overlaid on the image. The green-outlined shutters are those processed by the NIRSpec GTO pipeline for background subtraction and target spectrum extraction. The red-outlined shutter is excluded from the background subtraction process due to contamination of its trace by light from a non-target galaxy.
  }\label{f.contaminant}
\end{figure}

Fig.~\ref{f.pipeline} illustrates the NIRSpec GTO pipeline workflow, highlighting both the original NIPS steps and the updated modules developed by the GTO team. The pipeline processes the 2D count-rate products delivered by the MAST archive and treats each 2D count-rate image corresponding to each integration as an independent exposure. Another key input is the observed target catalogue, which includes the position of the targets in the MSA array and their locations within the shutter pitch. Additionally, the pipeline requires a supplemental catalogue, labelled “contaminated shutters” in Fig.~\ref{f.pipeline}, which lists all open MSA shutters that may contribute spectral traces to the detector due to the presence of an emitting source in those open shutters. As described below, this catalogue is crucial for background subtraction. It is generated by overlaying the open shutter positions onto available NIRCam or HST images and identifying overlaps with galaxies in the  open shutter aperture (Figure~\ref{f.contaminant}).

The pipeline executes eight default processing steps:
\begin{enumerate}
\item \emph{Pixel-to-pixel background subtraction}. For each target, the background emission is subtracted directly from the count-rate image using one or two shutters of the assigned slitlet that are free from contamination by other galaxies or extended emission from the target itself. This step is optional and may be skipped. As discussed in the next step, non-background-subtracted data can be used to identify problematic shutters and to create a global master background spectrum.

\item \emph{Extraction of the trace and assignment of the world coordinate system}. The pipeline extracts sub-images for each target, containing the spectral trace across three shutters, and assigns wavelength and spatial coordinates to each pixel. Spectral traces are extracted beyond the nominal wavelength range specified in the NIRSpec documentation for both R100 and R1000 observations, where there is still non-negligible throughput of the instrument. The wavelength ranges for each spectral configuration are summarized in Table~\ref{tab:wlrange} and shown in Fig. \ref{f.rosetta}.

\begin{table}
\begin{center}
\caption{Nominal and JADES  wavelength ranges of the disperser-filter combinations.}
\label{tab:wlrange}
\begin{tabular}{lccc}
  \hline
Disperser-filter &  Nominal range &  JADES range \\
  \hline
PRISM/CLEAR &0.6-5.3 & 0.6-5.45 \\ 
G140M/F070LP & 0.70–1.27 & 0.70-2.20\\
G235M/F170LP  &	1.66–3.07  & 1.66-4.00\\
G395M/F290LP  &	2.87–5.10  & 2.87–5.48 \\ 
G395H/F290LP&	2.87–5.14 &2.87–5.14\\
  \hline

\end{tabular}
\end{center}
\end{table}

\begin{figure*}
  \includegraphics[width=0.8\paperwidth]{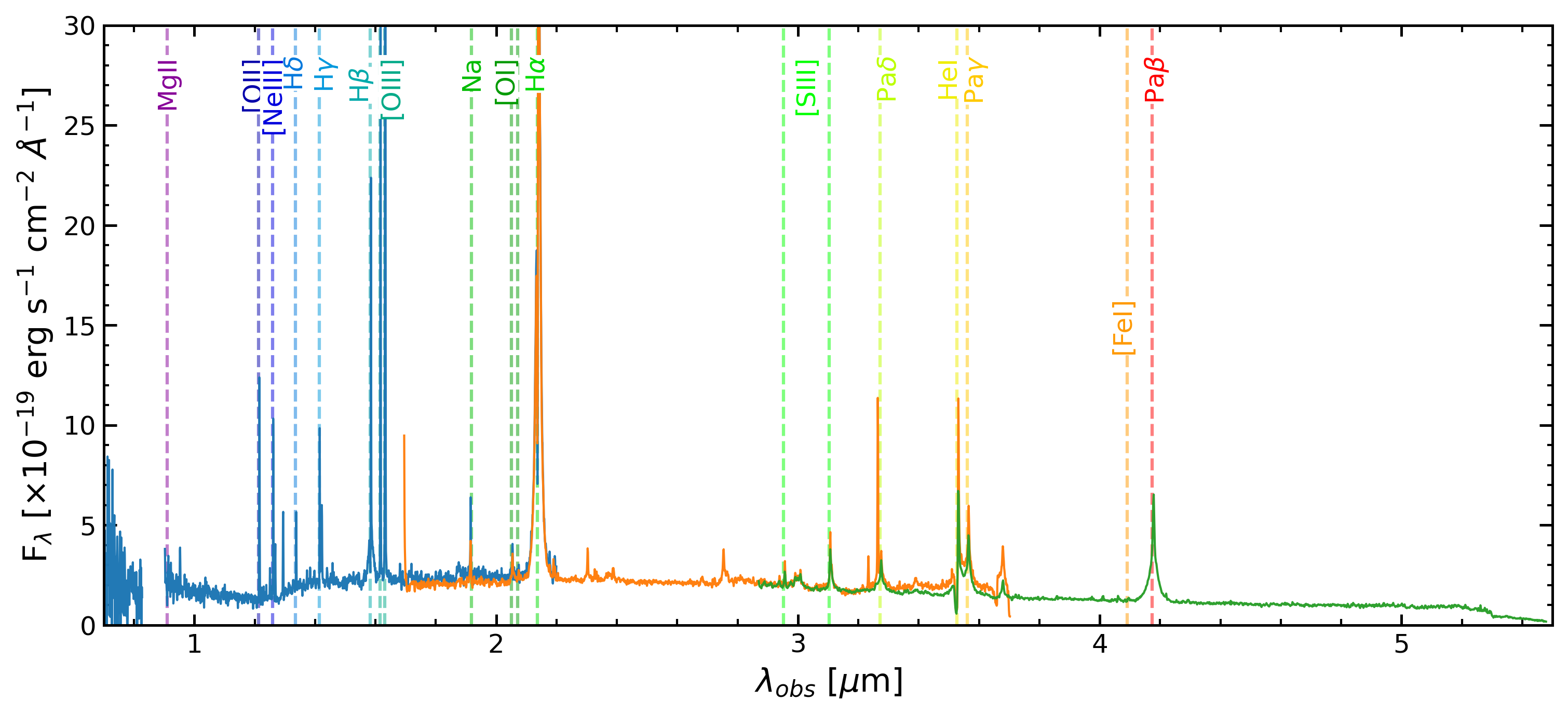}
  \caption{Illustration for the expanded wavelength coverage of gratings and the agreement in the flux calibration between the gratings of $<10\%$, even in the extended wavelength range. This object, \texttt{NIRSpec\_ID} 28074 from \texttt{goods-n-mediumhst} has been dubbed the Rosetta Stone of the Little Red Dots \citep[][]{Juodzbalis+2024_rosetta}.
  }\label{f.rosetta}
\end{figure*}

\item \emph{Wavelength correction}. A wavelength zero-point correction is applied to account for the intra-shutter position of each target. The correction map is provided in the WAVECORR reference file, which is based on Fourier optical simulations carried out prior to the JWST launch \citep{jakobsen+2022}. However, even after applying this correction, a wavelength offset discrepancy remains between emission lines measured in spectra obtained with the R1000 and R100 modes in previous data releases \citepalias[e.g.,][]{JADES_DR3} and in other surveys \citep[e.g.,][]{degraaff+2025}. We investigated this wavelength offset in detail and identified a correlation with both the wavelength and the intra-shutter position of the target along the dispersion direction. Based on this, we derived an additional, data-driven correction map that reduces the discrepancy between data obtained with the grating and the prism. This additional correction has been applied to the current data release. Further details of this analysis are provided in Section~\ref{s.wavecal}.

\item \emph{Flat-field correction}. The un-rectified trace of each target is corrected using the detector response curves called D-flat, S-flat, and F-flat curves \citep{rawle+2016}.  The D-flat characterizes how each pixel responds to different wavelengths, incorporating both quantum efficiency and the spectral performance of the anti-reflection coating. The F-flat accounts for throughput losses caused by reflections within the telescope and instrument optics, as well as the transmission properties of the selected filter. The S-flat describes how variations in aperture size and position affect the optical path of the light as it passes through different instrument components. For this data release, the Calibration Reference Data System (CRDS) context jwst1413.pmap was used. However, as mentioned above, we have extended the calibration beyond the nominal wavelength ranges of R1000 configurations to fully exploit the wavelength coverage of the spectrograph (see Table~\ref{tab:wlrange}). Appendix~\ref{app_ext} discusses how the transmission curve of each individual flat field reference file was extrapolated beyond the nominal wavelength range. 

\item \emph{Path loss correction}. A correction is applied for path losses, accounting for the deviation from a perfectly centered point source. This wavelength dependant correction assumes a point-like source geometry for all targets and uses the intra-shutter position of the target to determine the correction. Because the MAST archive includes observations of individual stars (PID 1133), we estimated the pathloss correction map directly from the data. This map provides correction factors as functions of both source position within the aperture and wavelength (see details in Appendix~\ref{app_pthl}).

 \item \emph{Resampling onto an irregular 2D grid}. This is one of the first modules modified by the NIRSpec GTO team members. In this step the extracted spectral traces are rectified onto a two-dimensional grid using an algorithm  by \cite{dorner+2012}, which is similar to  `drizzle' \citep{Fruchter2002}. The only distinction between the two algorithms is in the method used to estimate the variance. In the standard drizzle algorithm, the equation used to compute the variance includes the `fractional pixel overlap' squared (Equation 7 in \citealt{Fruchter2002}). Although this is mathematically correct, \cite{Fruchter2002} noted that this approach underestimates noise on larger spatial scales because it neglects the covariance terms introduced when data from a single input pixel are distributed over multiple output pixels. To compensate for this effect, the variance equation includes the fractional pixel overlap in linear form rather than squared. This results in an increase of approximately 15\% in the uncertainty for each individual output pixel but preserves the total noise level when integrating over multiple output pixels \citep{dorner+2012}. In the context of our final 1D spectra, the uncertainty is accurate only when estimated over a number of spectral channels corresponding to the line spread function and not at the level of individual spectral channels. A more precise noise estimate would require computation of the full covariance matrix. However, this is only feasible for those targets observed in more than 20 exposures, as detailed in \cite{hainline+2024b, carniani+2024} and \cite{witstok+2025}.

\item \emph{Extraction of master background spectrum}.  If the background subtraction step is skipped, the user can opt to have the pipeline generate a master background 2D image. This is done by computing the median of the 2D rectified, calibrated traces from all three-shutter slitlets designated as ``sky." The resulting master background image can then be subtracted from the 2D spectrum of the target to remove background emission. Based on our experience in the JADES program, background removal using the master background image is generally less accurate than the standard pixel-to-pixel subtraction method, particularly for compact galaxies. However, the master background approach can be advantageous for bright, extended galaxies, where diffuse emission contaminates all shutters of the slitlet assigned to the target. This step has been included in the NIRSpec GTO pipeline workflow, but is not yet implemented in this data release.

\item \emph{Extraction of irregular 1D spectrum}
In this step, the pipeline extracts 1D spectra from the rectified 2D maps. The 1D extraction is performed using boxcar apertures centered at the intra-shutter positions specified in the input catalogue. By default, a 5-pixel-wide aperture is used, corresponding to the width of a single open shutter. Additional extraction widths of 3 and 15 pixels are also available. A 3-pixel-wide aperture is optimal for compact sources and to optimise the signal-to-noise ratio of the final products. The 15-pixel-wide aperture should be used for extended galaxy with a size larger than 1 arcsec, corresponding to about 1.5 shutter (see \citealt{maseda+2024}). 

The extraction process accounts for the data quality array, similar to the procedure adopted in the STScI pipeline. Specifically, pixels flagged as  \texttt{DO NOT USE}, \texttt{SATURATED}, \texttt{OUTLIER}, \texttt{PERSISTENCE}, \texttt{DEAD}, \texttt{HOT}, and \texttt{OTHER BAD PIXEL} are excluded from the extraction.

If the background subtraction step is skipped, the individual PRISM/CLEAR 1D spectra are analyzed to identify unexpected closed shutters. The pipeline measures flux levels in the non-background-subtracted 1D spectra and flagged as ``disobedient'' those shutters which failed to open resulting in 1D spectra with a zero emission level.

\item \emph{Combination of 2D and 1D Spectra}. In this final step, the pipeline combines all available 2D and 1D spectra for each target, excluding exposures affected by disobedient shutters. The data are combined using inverse-variance weighting and the data quality array. For 2D spectra, alignment is performed based on the intra-shutter positions listed in the input catalogue.

Prior to combination, the pipeline masks pixels with significantly negative flux values as these are considered unphysical and typically arise from outliers or serendipitous self-subtracted emission in the slitlets. A sigma-clipping algorithm is then applied to remove remaining outliers. This algorithm flags as outliers any pixels that deviate by more than three standard deviations from the mean value at each wavelength channel. However, this algorithm generally requires a minimum of six exposures per target to perform an appropriate outlier rejection. The pipeline, therefore, provides two versions of the final combined spectra: one with and one without sigma clipping, allowing users to assess data quality.

It is important to stress that the final 1D spectrum is not extracted from the combined 2D map. Instead, it results from the combination of individual 1D spectra. This approach is more rigorous as it avoids inaccuracies that may arise from varying intra-shutter positions across different visits, which can alter the cross-dispersion surface brightness profile due to path-loss effects.

If more than 20 exposures are available for a given target, the covariance matrix of the combined spectrum is computed from 2,000 bootstrap realizations, following the methodology described in \cite{hainline+2024b} and \cite{witstok+2025}. The full details of this algorithm and a comprehensive analysis of correlated noise in the final spectra will be presented in Jakobsen et al. in preparation.

\end{enumerate}

For this data release, we applied the NIRSpec GTO pipeline three times by default for PRISM/CLEAR observations. First, we processed the data without background subtraction in order to identify disobedient shutters. Next, the pipeline was run with standard background subtraction, extracting 1D spectra using both 5-pixel and 3-pixel apertures. Finally, we reprocessed the data considering only exposures from the top and bottom shutters of the three-shutter slitlets, excluding those where the target was positioned in the central shutter. This additional processing helps mitigate self-subtraction effects in extended galaxies during background removal. In the case of observations using two nod positions, we also performed spectral extractions with both 3-pixel and 5-pixel aperture widths in the cross-dispersion direction. For the R1000 and R2700 observations, the pipeline was applied following the same approach used for the PRISM/CLEAR data, with the exception of the initial run without background subtraction. Instead, the list of disobedient shutters identified from the R100 data was adopted for processing the higher-resolution observations.

\subsection{GTO pipeline vs JWST Science Calibration Pipeline}

Most of the steps and algorithms of  the NIRSpec GTO pipeline are similar to those encapsulated within the official JWST Science Calibration Pipeline. However, there are some differences in the workflow that might improve the quality of the final products: 
\begin{itemize}

\item \emph{Identification of contaminants in the 3-shutter slitlet}. This step improves pixel-by-pixel background subtraction by identifying and removing off-scene nods that are contaminated by galaxies near the targets that fall in one of the free-target shutter. Without this correction, light from these galaxies would be included in the predicted background emission and subtracted from the spectrum of the target, artificially reducing its observed emission.

\item \emph{New wavelength correction based on the position of the target in the shutter.} The current version of the JWST Science Calibration Pipeline performs a wavelength zero-point correction for an off-centre target, but the model used by this step is not sufficient to calibrate the wavelength array and this causes a spectral offset between grating and prism data. Indeed, MAST final products reveal a systematic spectral offset between the line centroids measured from the prism spectra and those measured in the grating data (Figure~\ref{f.app_mastwavecal}). This spectral offset is not constant but depends on the location of the target in the MSA shutter, the NIRSpec GTO pipeline adopts a new data-based model to perform the wavelength calibration which corrects the systematic spectral offset between grating and prism.

\item \emph{Identification of disobedient shutters.} This step identifies shutters that are unexpectedly closed. Such disobedient shutters produce noisy 1D and 2D spectra lacking both background and target emission. If not excluded during the combination process, they can corrupt the final spectra. The NIRSpec GTO pipeline automatically detects these failed shutters and removes them from the data processing workflow.

\end{itemize}

\subsection{High Level Science Products}\label{s.hlsp}

The full suite of reduced data products for the JADES programs is available at MAST as a High Level Science Products (HLSP) via \url{https://archive.stsci.edu/hlsp/jades}, \url{https://mast.stsci.edu/hlsp}, and \url{https://jades-survey.github.io}.
This release provides reduced 1D and 2D spectra with multiple options for data processing and 1D spectral extraction. Specifically, the data are processed in two ways: (1) using all three nods for each target (default workflow), and (2) using only the exposures from the top and bottom microshutters (``2 nods'') for both the pixel-by-pixel background subtraction and the combination process. In the context of the extraction of 1D spectra, we use the default full shutter aperture (``5 pix'') and the small box-car aperture of 3 pixels (``3 pix'') centred at the target's location.

The 1D and 2D spectra  are available for each spectral configuration and field reported in Table~\ref{tab:obssummary} according to the following convention:
\begin{itemize}
\item \emph{hlsp\_jades\_jwst\_nirspec\_<TIER>-<NIRSpec\_ID>\_<filter>-<grating>\_v1.0-spec1d.fits}
\item \emph{hlsp\_jades\_jwst\_nirspec\_<TIER>-<NIRSpec\_ID>\_<filter>-<grating>\_v1.0-spec2d.fits}
\end{itemize}
where {\it <TIER>} and {\it <NIRSpec\_ID>} are the name of the tier and the ID of the NIRSpec target, respectively. {\it <filter> } and {\it <grating>} refer to the disperser-filter combination used in the observations. The fits files of the 1D PRISM/CLEAR spectra have this structure:
\\
\vspace{1pt}
\begin{tabular}{lccc}
No. &  Name &  Type & Columns \\
0   &  \texttt{PRIMARY} & PrimaryHDU   \\ 
1   &  \texttt{EXTRACT5PIX1D} & BinTableHDU   \\ 
2   &  \texttt{EXTRACT3PIX1D} & BinTableHDU \\
3   &	\texttt{INTERMEDIATE}  & BinTableHDU\\
4   &	\texttt{REGION}  & BinTableHDU \\ 
5   &	\texttt{F090W} & Image\\
6   &	\texttt{F150W} & Image\\
7   &	\texttt{F277W} & Image\\
8   &	\texttt{F444W} & Image\\
\vspace{1pt}
\end{tabular}\\
where \texttt{EXTRACT5PIX1D} and \texttt{EXTRACT3PIX1D} products provide the ``5 pix'' and ``3 pix'' 1D spectra extracted from the 2D maps, respectively. Each table includes both the spectra obtained with the standard ``3-nod'' scheme and those derived using only the ``2-nod'' exposures. For the default nodding scheme, we also report the ``5 pix'' 1D spectra extracted from all individual exposures in the \texttt{INTERMEDIATE} extension. The fits file also includes 4 wide-band flux-calibrated NIRCam images of 5arcsec$\times$5arcsec around the target (see Johnson et al., in prep for details of the NIRCam reduction). Finally, the extension \texttt{REGION} is a FITS Region binary table reporting  the corner positions of the slitlets used for the target in pixel units, which can be used directly in the attached NIRCam images, and the respective RA and Dec in the International Celestial Reference System (ICRS)\footnote{We note that the astrometric solution for GOODS-N changed between target selection and final NIRCam reductions (see Johnson et al., in prep).  We include an average, per-object shift to the shutter regions but caution that there is an additional level of uncertainty in their exact positions due to these changes.}. We note that this extension is supported by \texttt{DS9}. The 1D HLSP related to the observations obtained with the grating have the same structure but they do not include the NIRCam cutout images.

The 2D HLSP fits files have this structure:
\\
\begin{tabular}{lccc}
No. &  Name &  Type  \\
0   &  \texttt{PRIMARY} & Image   \\ 
1   &  \texttt{FLUX} & Image   \\ 
2   &  \texttt{FLUX\_ERR} & Image \\
3   &	\texttt{WAVELENGTH}  & Image\\
4   &	\texttt{FLUX\_2NOD}  & Image \\ 
5   &	\texttt{FLUX\_ERR\_2NOD} & Image\\
\vspace{1pt}
\end{tabular}\\
where \texttt{FLUX} and \texttt{FLUX\_ERR} illustrate the combined flux-calibrated 2D maps from the standard ``3 nod'' scheme, while the 2D maps for the ``2 nod'' data processing workflow are included in the extension \texttt{FLUX\_2NOD} and \texttt{FLUX\_ERR\_2NOD}. Finally, the wavelength array is reported in \texttt{WAVELENGTH}.

\section{prism emission-line fluxes}\label{s.r100}

The method used to fit the PRISM spectra is the same as outlined in \citetalias{JADES_DR3}. In order to fit the PRISM spectra with the underlying strong continuum, we use the spectral fitting software \ppxf \citep{cappellari2023}, which models the data as a linear combination of physically motivated spectral templates to ensure that the underlying stellar absorption (in particular Balmer absorption) is accounted for in the emission line fluxes. As input templates, we considered a set of simple stellar population templates from \fsps \citep[SSP;][]{conroy+2009}. The templates were calculated using \textsc{mist} isochrones \citep{choi+2016}, the C3K model atmospheres \citep{conroy+2019}, and a Salpeter initial mass function \citep{salpeter1955}. The templates have higher resolution (R=10,000) between $0.1<\lambda<3$~\mum; compared to the PRISM resolution of R=50--300. The templates are available from C. Conroy upon reasonable request. To reduce degeneracies, we only consider a subset of templates, which spans ages in the range of $0.03\text{--}20$~Gyr and metallicities [Z/H] $-2.5\text{--}0$ in logspace. For each target, we also adjust the grids to ensure that the oldest available SSP is consistent with the age of the Universe at the redshift of the fitted target, save a buffer of one age bin \citep[see e.g.,][]{looser+2023}.

In addition to the stellar templates, we also fit a set of Gaussian models to represent the nebular emission lines in the data. These nebular/gas templates can be divided into three separate categories: i) single Gaussian components representing individual emission lines (or doublets/multiplets that are completely unresolved)  that are spectrally isolated at any redshift at the resolution of the PRISM observations (e.g., \HeIL[5787], \Pabeta); ii) single Gaussian component representing the spectrally blended lines given the resolution of PRISM (e.g., $\Halpha+\NIIall$, $\Hgamma+\OIIIL[4363]$); and iii) two Gaussian components representing doublets with fixed ratios (e.g., \OIIIall, \SIIIall). We summarise the emission line templates and their redshift range in Table~\ref{t.templates}. We note that the exact set of templates that we use in the fitting is dependent on the sources' initial redshift, since the spectral resolution of the prism is strongly evolving with wavelength, making some emission-line groups spectrally resolved or unresolved at different redshifts. Following \citet{cameron+2023c} and \citetalias{JADES_DR3}, we also include a step function that is designed to capture the strong Balmer jumps that appear in some of our objects. 

We set bounds on the linear combination coefficients to be always positive. Finally, we also use a 10\textsuperscript{th}-order multiplicative Legendre polynomial to adapt the shape of the continuum to the data; this can be thought of as a combination of physical effects (e.g., dust reddening) and flux calibration issues (e.g., incorrect slit-loss corrections, for extended objects and for objects with strongly wavelength-dependent morphology, flat fielding effects, etc..). Before we start the \ppxf run, each template is convolved to match
the
spectral resolution of the data; we use the nominal resolution, boosted by a factor of 1/0.7.
The reason for increased effective resolution
is that the nominal resolution is given for the case of uniformly illuminated shutter, while in the case of distant galaxies (especially those at very high redshifts) they are typically smaller than the shutter width, hence their size sets the effective resolution \citep{degraaff+2023}.
A similar resolution boost has been adopted independently by other works \citep[e.g.,][]{Greene+2024}. We remark that we do not use the derived line widths from the prism. We truncate the templates to match the rest-frame wavelength range of our data. The stellar templates are set to zero blueward of \Lyalpha, i.e. we do not include this region in the fit. Finally, we convolved the templates with an instrumental line spread function, which is modelled as a Gaussian shape. 

Overall, we run our \ppxf fitting twice for each galaxy. During the first fit, we fixed the templates' kinematics subsets, constraining them to have the same central velocity and velocity dispersion. The kinematic groups are: Balmer lines and stellar templates, rest-frame UV lines, rest-frame optical lines and rest-frame NIR lines. After the initial round of fitting, we discard any emission lines that are not detected at least  5\textsigma significance. 

During the second run, we fix the kinematics of the stellar continuum absorption, use only previously detected emission line templates, and remove almost all kinematic groups. Exceptions to the latter rule are: the blend group formed by \SIIall and the blend $\Halpha+\NIIall$; the group formed by \Hbeta and \OIIIall; the group formed by \Hgamma and \OIIIL[4363]; and the group of \HeIL[10830] and \Pagamma, whose kinematics are always tied together. It is important to note, that \HeIL[10830] is resonant, therefore, this emission-line tends to be redshifted relative to the systemic velocity of the target; however, leaving the line kinematics free relative to \Pagamma tended to produce bad fits due to low spectral resolution. Therefore, we opted to keep these lines tied. This setup closely follows the one outlined in \citetalias{JADES_DR3}, as it is necessary due to the limited spectral resolution of the prism observations. 

After each fit, we post-process the line fluxes as follows. Below redshift $z<2$, we combine \Hbeta and \OIIIall and $\Halpha+\NIIall$ and \SIIall; the line uncertainties are added in quadrature.
Between $2\leq z < 5.3$, we combine the flux from the \OIIIall doublet. The best-fit spectra are carefully inspected for any artefacts or failed fits, either due to low equivalent width emission lines near the Balmer break, emission lines due to contaminants, and outliers, etc. The low equivalent width emission lines near the Balmer break arise when the shape of the break is not fit correctly, and the algorithm may use \OIIall, \OIIIall, \NeIIIall and \Hdelta to add to the continuum. Contaminants and artifacts may escape the sigma-clipping in \ppxf when they fall close to strong emission lines in the intended target.

\begin{table*}
\begin{center}
\caption{List of the emission lines fit in the prism spectra. All wavelengths are in vacuum.}\label{t.templates}
\begin{tabular}{ll|lcll}
    \hline
    & Line(s) & $\lambda$ [\AA] & $z$ range & Column name & Notes \\ 
    \hline
    & \CIVall                 & 1549.48         &      ---     & \verb|C4_1549|          & \\
    & \HeIIL+\semiOIIIall     & 1650.00         &      ---     & \verb|Blnd_He2_O3_1650| & \\
    & \CIIIall                & 1907.71         &      ---     & \verb|C3_1907| & \\
    & \MgIIall                & 2799.94         &      ---     & \verb|Mg2_2796| & \\
    & \OIIall                 & 3728.49         &      ---     & \verb|O2_3727| & \\
    & \NeIIIall               & 3869.86,3968.59 & $0 <z < 5.3$ & \verb|Ne3_3869|,\verb|Ne3_3968| & \\
    & \NeIIIL[3869]           & 3869.86         & $z \geq 5.3$ & \verb|Ne3_3869| & \\
    & \NeIIIL[3968]+\Hepsilon & 3968.59         & $z \geq 5.3$ & \verb|Ne3_3968|      & \\
    & \Hdelta                 & 4102.86         &      ---     & \verb|HD_4102| & \\
    & \Hgamma+\OIIIL[4363]    & 4341.65         & $0 <z < 5.3$ & \verb|Blnd_HG_O3 | & \\
\rdelim\{{2}{*}[] & \Hgamma                 & 4341.65         & $z \geq 5.3$ & \verb|HG_4341|  & \\
    & \OIIIL[4363]            & 4363.44         & $z \geq 5.3$ & \verb|O3_4363|  & \\
\rdelim\{{2}{*}[] & \Hbeta    & 4862.64         & $0 < z < 2 $ & \multirow{2}{*}{\texttt{Blnd\_HB\_O35007d}} & \\
    & \OIIIall                & 4960.30,5008.24 & $0 < z < 2 $ & & \\
\rdelim\{{2}{*}[] & \Hbeta    & 4862.64         & $2 \leq z < 5.3$ & \verb|HB_4861| & \\
    & \OIIIall                & 4960.30,5008.24 & $2 \leq z < 5.3$ & \verb|O3_5007d| & \\
\rdelim\{{2}{*}[] & \Hbeta    & 4862.64         & $z \geq 5.3$ & \verb|HB_4861| & \\
    & \OIIIall                & 4960.30,5008.24 & $z \geq 5.3$ & \verb|O3_4959|,\verb|O3_5007| & \\
    & \HeIL[5875]             & 5877.25         &      ---     & \verb|He1_5875| & \\
    & \OIall                  & 6302.05,6363.67 &      ---     & \verb|O1_6300| & \\
\rdelim\{{2}{*}[] & \Halpha+\NIIall & 6564.52   & $0 < z < 2 $ &  \multirow{2}{*}{\texttt{Blnd\_HA\_N2\_S2}} & \\
    & \SIIall                 & 6725.00         & $0 < z < 2 $ & & \\
\rdelim\{{2}{*}[] & \Halpha+\NIIall & 6564.52   & $ z \geq 2 $ & \verb|HA_6563| & \\
    & \SIIall                 & 6725.00         & $ z \geq 2 $ & \verb|S2_6725| & \\
    & \HeIL[7065]             & 7067.14         &      ---     & \verb|He1_7065| & \\
    & \SIIIall                & 9071.10,9533.20 &      ---     & \verb|S3_9069|,\verb|S3_9532| & \\
    & \Padelta                & 10052.12        &      ---     & \verb|PaD_10049| & \\
\rdelim\{{2}{*}[] & \HeIL[10829]            & 10832.06        &      ---     & \verb|He1_10829| & \\
    & \Pagamma                & 10940.98        &      ---     & \verb|PaG_10938| & \\
    & \Pabeta                 & 12821.43        &      ---     & \verb|PaB_12818| & \\
    & \Paalpha                & 18755.80        &      ---     & \verb|PaA_18751| & \\
    \hline
\end{tabular}
\end{center}
The set of templates used to fit any given galaxy depends on its initial redshift guess; this is because the spectral resolution of the prism is a strong function of wavelength \citep{jakobsen+2022}, causing emission-line groups to be spectrally resolved or unresolved at different redshifts. Empty redshift ranges indicate the template is used at all redshifts.
Rows connected by curly braces indicate emission-line pairs/groups that have tied velocity and velocity dispersion.

\end{table*}
\begin{table}
\begin{center}
\caption{Structure of the prism flux table. The full list of emission lines is reported in Table~\ref{t.templates}; all fluxes are in units of \fluxcgs[-20].}\label{t.prism}
\begin{tabular}{ll}
    \hline
    Column name               & Description \\
    \hline
    \texttt{Unique\_ID}       & Unique ID of the source in the survey  \\
    \texttt{PID}              & Program ID \\
    \texttt{TIER}             & Name of subset$^\ddag$ \\
    \texttt{TIER\_old}         & Old Name of subset used in DR1 and DR3 \\
    \texttt{NIRSpec\_ID}      & ID of the target in eMPT$^\ddag$ \\
    \texttt{NIRCam\_DR5\_ID}  & NIRCam ID from upcoming DR5  \\
    \texttt{NIRCam\_DR3\_ID}  & NIRCam ID from DR3 (\citetalias{JADES_DR3})\\
    \texttt{z\_PRISM}         & Prism-based redshift                    \\
    \texttt{[name]\_flux}   & emission line flux             \\
    \texttt{[name]\_flux\_err}    & emission line flux uncertainty \\
    \hline
\end{tabular}
\end{center}
$^\ddag$ \texttt{NIRSpec\_ID} are not unique in the table, but the combination of 
\texttt{NIRSpec\_ID} and \texttt{TIER} is unique.
$^\dagger$ \texttt{NIRCam\_ID}s are unique, but whether they match the \texttt{NIRSpec\_ID}s depends on target selection (\hst vs \jwst selection), as well as on whether the NIRCam catalog was revised after the NIRSpec observation (which may result in sources being lost to blending and to crossing the non-detection threshold).
\end{table}

\section{Medium-resolution gratings emission-line fluxes}\label{s.r1000}

We fitted the medium-resolution spectra using  \texttt{QubeSpec}'s\footnote{\url{https://github.com/honzascholtz/Qubespec}} fitting module, specifically designed for NIRSpec/MSA observations. We fitted each emission line with a single Gaussian component and the continuum with a simple power law. Although this is a simplistic approach, it is perfectly adequate for describing a narrow range of the continuum around an emission line of interest ($\pm100 ~\AA$), as usually the continuum is poorly detected. The majority of the emission lines are fitted in isolation, except for a group of emission lines that are close to each other, such as Halpha, [NII] and [SII]. We show the full list of emission lines fitted in this work and the groups fitted together in Table \ref{grating.eml} along with names of the emission lines in our public release. 

\begin{table*}
\begin{center}
\caption{List of the emission lines fit in the medium-resolution grating spectra. All wavelengths are in vacuum. Rows connected by curly braces indicate emission lines that were fitted using the same redshift and FWHM during the same fit because they are sufficiently close in wavelength that the continuum can be modeled simultaneously. }\label{grating.eml}
\begin{tabular}{ll|ll}
    \hline
    & Line(s)                   & $\lambda$ [\AA]  & Column name \\ 
    \hline
    \rdelim\{{3}{*}[]& \CIVall  & 1549.48          & \verb|C4_1549|  \\
    & \HeIIL                    & 1640.00          & \verb|He2_1640| \\
    & \semiOIIIall              & 1663.00          & \verb|O3_1663|  \\
    & \CIIIall                  & 1907.71          & \verb|C3_1907|  \\
    \rdelim\{{2}{*}[] & \OIIall & 3728.49          & \verb|O2_3727|  \\
    & \NeIIIL[3869]             & 3869.86          & \verb|Ne3_3869| \\
    & \Hdelta                   & 4102.86          & \verb|HD_4102|  \\
\rdelim\{{2}{*}[] & \Hgamma     & 4341.65          & \verb|HG_4341|  \\
    & \OIIIL[4363]              & 4363.44          & \verb|O3_4363|   \\
\rdelim\{{2}{*}[] & \Hbeta      & 4862.64          & \verb|HB_4861| \\
    & \OIIIall                  & 4960.30,5008.24  & \verb|O3_5007| \\
    & \HeIL[5875]               & 5877.25          & \verb|He1_5875| \\
    & \OIall                    & 6302.05          & \verb|O1_6300| \\
\rdelim\{{3}{*}[] & \Halpha     & 6564.52          &  \verb|HA_6563| \\
    &   \NIIall                 & 6585.27, 6549.86 & \verb|N2_6584|   \\
    & \SIIall                   & 6718.29, 6732.67 & \verb|S2_6718|, \verb|S2_6732| \\
    & \HeIL[7065]               & 7067.14          & \verb|He1_7065| \\
    & \SIIIall                  & 9071.10,9533.20  & \verb|S3_9069|,\verb|S3_9532| \\
    & \Padelta                  & 10052.12         & \verb|PaD_10049| \\
    & \HeIL[10829]              & 10832.06         & \verb|He1_10829| \\
    & \Pagamma                  & 10940.98         & \verb|PaG_10938| \\
    & \Pabeta                   & 12821.43         & \verb|PaB_12818| \\
    & \Paalpha                  & 18755.80         & \verb|PaA_18751| \\
    \hline
\end{tabular}
\end{center}
\end{table*}

To estimate the model parameters, we use a Bayesian approach implemented with the Markov-Chain Monte-Carlo (MCMC) integrator \texttt{emcee} \citep{foreman-mackey+2013}. To measure the emission-line fluxes, redshifts and widths, we need to set prior probabilities for each of the variables. The peaks of the Gaussian profiles and the continuum normalization are given a log-uniform prior, while the FWHMs are set to a uniform distribution spanning from the minimum spectral resolution of the NIRSpec/MSA ($\sim$200 \kms) and 800~\kms. The prior on the redshift was a truncated normal distribution centered on the redshift from the visual inspection and with a standard deviation of 300~\kms and with a maximum allowed deviation of 1,000~\kms. 

We fit only a single Gaussian per emission line in the medium-resolution grating. We note that there are some objects with detected outflows or broad line regions. The potential type-1 AGN have been investigated in  \citep{juodzbalis+2025} and broad components in forbidden lines associated with outflows will be presented in  S.~Carniani et~al. (in~prep.).

After the initial fitting run, we visually inspect every model for any incorrect fits or spurious line detection that are caused by unflagged outliers or other problems. These flagged fits are then refitted and re-inspected to ensure sufficient quality. The final fluxes are calculated using the MCMC chains (after discarding the burn-in chains - 50\%) and the final reported values and their uncertainties are the median value and standard deviation from the chains. 
We estimated the SNR of an emission line as the 50\%/16\% of the posterior distribution of the emission line flux. This ensures we correctly estimate SNR of an emission line for lines with highly asymmetric posterior distribution of emission line flux. 
The final redshift from the medium-resolution spectra is the redshift inferred as a weighted average of emission lines with SNR$>$5.

The structure of the gratings emission line catalogue is presented in Table \ref{grating.eml_structure} where the fluxes are in units of $\times 10^{-20}$ erg s$^{-1}$ cm$^{-2}$. The names of the individual emission line columns are the same as reported in Table \ref{grating.eml}. For the emission line doublets with fixed line ratios (such as \NIIall and \OIIIall) we only report the flux of the stronger emission line. Alongside the fluxes and their uncertainties, the initial rows are the same as for the prism table. 

Given the spectral overlap of the gratings, some emission lines are measured in two adjacent gratings. For emission lines that lie in the spectral range where the gratings overlap, we fit both sets of data.  We do not attempt to stack these spectral overlaps due to different line spread functions and potential flux calibration offsets between the gratings that can bring further uncertainties into the fit. The final emission line measurements are primarily chosen from the redder of the available disperser, unless the emission line is not covered by the instrument due to the detector. We include a column in the final R1000 table (\texttt{eml\_name\_filter}) indicating the grating/filter combination used to estimate the value in the table. Furthermore, we also add \texttt{ext} to the name of the disperser/filter combination to indicate when the emission line was extracted outside of the nominal spectral range of NIRSpec disperser/gratings presented in Table \ref{tab:wlrange}.

\begin{table}
\begin{center}
\caption{Structure of the gratings flux table. The initial rows are the same as for the prism
(between \texttt{NIRSpec\_ID} and \texttt{z\_PRISM}; cf.~Table~\ref{t.prism}); all fluxes are in units of \fluxcgs[-18].
}\label{grating.eml_structure}
\begin{tabular}{ll}
    \hline
    Column name               & Description \\
    \hline
    \texttt{Unique\_ID}       & Unique ID of the source in the survey  \\
    \texttt{PID}              & Program ID \\
    \texttt{TIER}             & Name of subset$^\ddag$ \\
    \texttt{TIER\_old}         & Old Name of subset used in DR1 and DR3 \\
    \texttt{NIRSpec\_ID}      & ID of the target in eMPT$^\ddag$ \\
    \texttt{NIRCam\_DR5\_ID}  & NIRCam ID from upcoming DR5  \\
    \texttt{NIRCam\_DR3\_ID}  & NIRCam ID from DR3 (\citetalias{JADES_DR3})\\
    \texttt{z\_R1000}         & R1000-based redshift                    \\
    \texttt{z\_R1000n}        & Number of emission lines used to determine\\                                      &  the best R1000 redshift      \\
    \texttt{[name]\_flux}      & emission line flux             \\
    \texttt{[name]\_err}       & emission line flux uncertainty \\
    \texttt{[name]\_SNR}       & SNR emission line based on 50\%/16\% \\
    & percentile of the posterior distribution  \\
    \texttt{[name]\_filter}    & emission line disperser/filter  \\
    \hline
\end{tabular}
\end{center}
$^\ddag$ \texttt{NIRSpec\_ID} are not unique in the table, but the combination of 
\texttt{NIRSpec\_ID} and \texttt{TIER} is unique.
\end{table}

\section{Quality assessment}\label{s.quality}

In this section, we perform the quality assessment of the data reduction and the data analysis. In \S~\ref{s.wavecal} we describe the accuracy of the wavelength calibration, in \S~\ref{s.flux_cal} \& \ref{s.flux_cal_inter} we describe the absolute flux calibration of our data products along with the comparison of the prism and gratings observations, and in \S~\ref{s.redshift_quality} we describe the quality of our redshift estimates. 

\subsection{Accuracy of the wavelength calibration}\label{s.wavecal}

\begin{figure}
  \includegraphics[width=0.95\columnwidth]{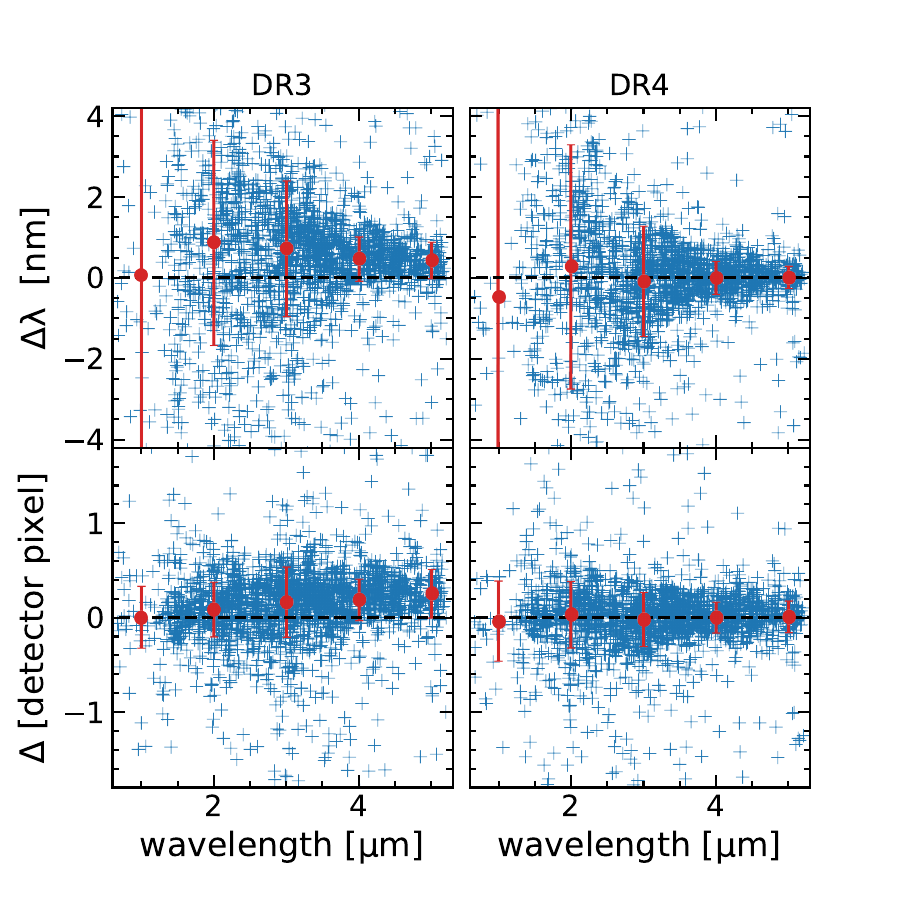}
  \caption{Comparison of the line centroids of \OIII, \Halpha, and \HeI measured in the prism and grating data for DR3 (left) and DR4 (right). The top panels show the wavelength offset in units of nm, while the bottom panels report the spectral offset in units of detector native prism pixels. Red marks show the median and standard deviation values of the distribution of measurements at 1, 2, 3, 4, and 5 $\mu$m.
  }\label{f.wloffset}
\end{figure}

Several \jwst studies have reported a discrepancy between redshifts measured from prism and grating data for the same targets \citep[e.g.][]{bunker+2023a, JADES_DR3, degraaff+2025}. To investigate the origin of this discrepancy and to identify a possible empirical solution, we exploited the extensive JADES dataset. Initially, we processed the data using the same pipeline employed for DR3, but updated with the latest standard reference files from the Calibration Reference Data System (CRDS; see Section~\ref{sec:observations}). We then measured the spectral centroids of emission lines in both the prism and grating data, focusing on the brightest lines: \OIII, \Halpha, and \HeI. In the right panels of Figure~\ref{f.wloffset} we show the offset between the two observational modes, expressed both in wavelength and in native detector pixel units. We found a systematic offset ranging from 0.5 to 2.1 nm, corresponding to approximately 0.15 to 0.3 detector pixels for the prism. The variation of the offset in wavelength units is strongly wavelength-dependent, primarily due to the non-uniform spectral resolution of the prism across its wavelength range. For this reason, all further analysis was carried out in detector pixel units to minimise the impact of this dependence.

Our analysis revealed that the spectral offset correlates with the intra-shutter position of the target along the dispersion direction. A similar result was also reported by \citetalias{JADES_DR3} and \cite{degraaff+2025}. This suggests that the centroid of the spectral point spread function, which depends on the location of the source within the shutter, is not fully accounted for in the current calibration. In particular, we suspect that the “wavecorr” step, which is intended to update the wavelength solution based on the offset of the source from the shutter center, does not apply a sufficiently accurate correction.

We therefore calculated the additional wavelength shifts ($S(\lambda,x)$) to apply in addition to the shift predicted by the CRDS reference file $S^{ref}(\lambda,x)$. In detail, the correction can be expressed as 
\begin{equation}\label{eq.wlzpcorr}
\lambda^{corr} = \lambda+[S(\lambda,x)+S^{ref}(\lambda,x)]\delta\lambda(\lambda)
\end{equation}
where $ \lambda$ is the wavelength assigned to the detector pixel for a point source at the centre of a specific shutter of the MSA array and $\delta\lambda(\lambda)$ is the size in wavelength unit of the detector pixel. Therefore, $S(\lambda,x)$ can be estimated assuming that the centroids of the emission lines of the prism and grating are identical after applying the additional correction:
$$
0=\lambda_{prism}^{corr}-\lambda_{grat}^{corr} 
$$
$$
= \lambda_{prism}-\lambda_{grat}+[S(\lambda,x)+S^{ref}(\lambda,x)][(\delta\lambda_{prism}(\lambda)-\delta\lambda_{grat}(\lambda)]
$$
Since $\delta\lambda_{prism}(\lambda)$ and $\delta\lambda_{grat}(\lambda)$ are known and the measured wavelength offset reported in Figure~\ref{f.wloffset} corresponds to:
$$
\Delta \lambda(\lambda,x) = \lambda_{prism}-\lambda_{grat}+S^{ref}(\lambda,x)[\delta\lambda_{prism}(\lambda)-\delta\lambda_{grat}(\lambda)]
$$
we can map the additional $S(\lambda,x)$ correction at varying $\lambda$ and intra-shutter position $x$ as:
$$
S(\lambda,x)= - \frac{\Delta \lambda (\lambda,x)}{\delta\lambda_{prism}(\lambda)-\delta\lambda_{grat}(\lambda)}
$$

\begin{figure}
  \includegraphics[width=0.95\columnwidth]{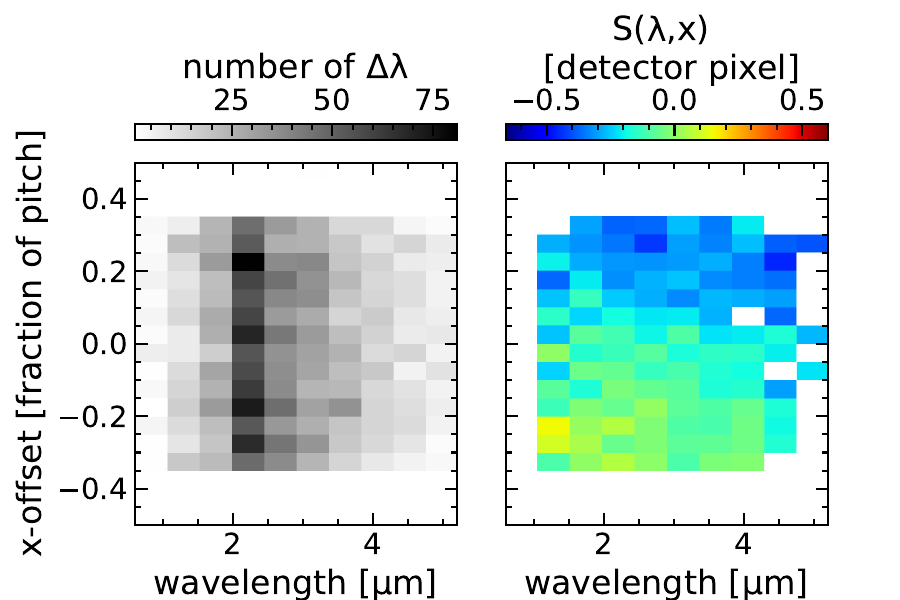}
  \caption{Empirical wavelength zero-point additional offset for a point-source as a function of
wavelength and source displacement from the micro-shutter centre in the
dispersion direction. Left panel shows the distribution of measurements while the right panel reports the offset to apply on the map in addition to the predictions from the CRDS reference file.}\label{f.shift}
\end{figure}

The outcome of this analysis is illustrated in Figure~\ref{f.shift} which reports the distribution of the number of spectral offset measurements $\Delta \lambda$ and a discrete map of $S(\lambda,x)$. The grid of the map was selected to maximise the coverage of the shutter aperture and have at least 10 measurements for each bin. 

The empirical correction map $S(\lambda, x)$ was incorporated into the data-processing workflow by applying linear interpolation onto the finer grid adopted by the pipeline, and linear extrapolation at wavelengths or intra-shutter offsets not covered by our data. The interpolated map was then applied to process the spectra used in this analysis. As shown in the right panels of Fig.~\ref{f.wloffset}, the systematic offset between line centroids measured in the prism and those measured in the grating is significantly reduced. However, we note a substantial scatter in the distribution of measurements, particularly at shorter wavelengths. This scatter may be attributed to both the uncertainty in centroid measurements in the prism spectra, especially at $1~\mu$m, where the spectral resolution is low, and residual inaccuracies in the wavelength calibration associated with the spatial extension of the targets and uncertainties on the target location within the shutter.

\begin{figure}
  \includegraphics[width=0.95\columnwidth]{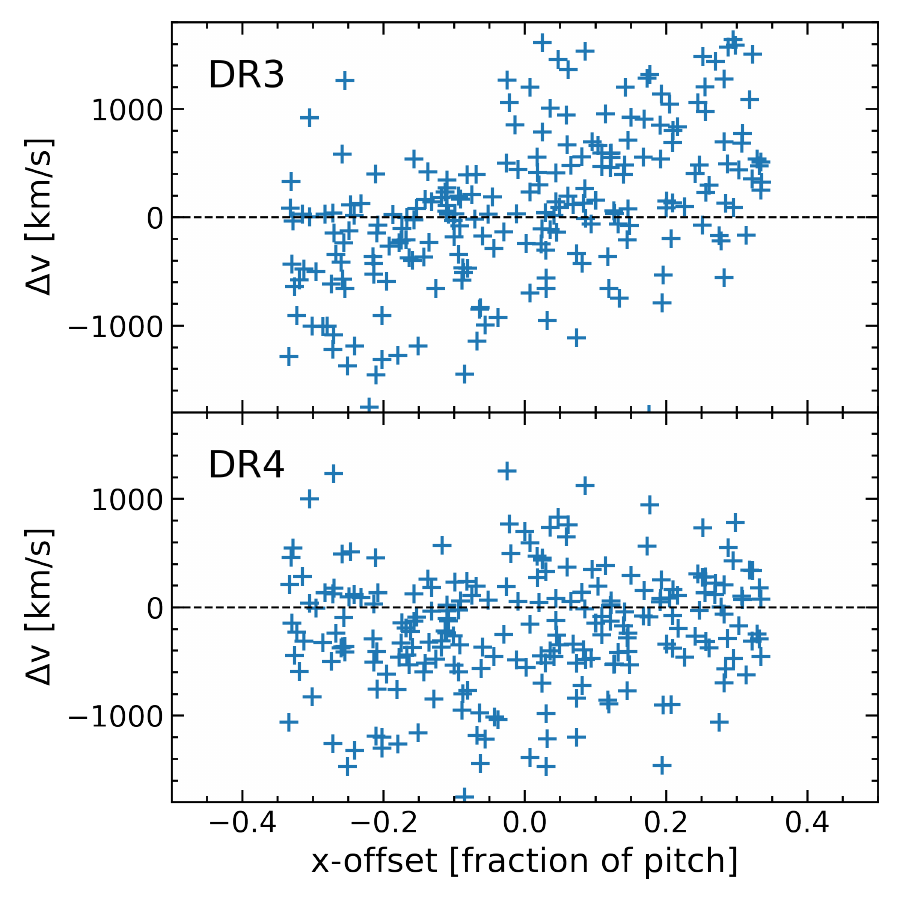}
 \caption{Wavelength calibration bias as a function of intrashutter source position for prism. $\Delta v$ indicates the wavelength calibration bias as defined in equation~\ref{eq.prismcal}, x-offset is the spatial offset of the target with respect to the centre of the shutter, measured along the dispersion direction. For an unbiased wavelength solution, we would expect  $\Delta v  = 0$ at any x-offset values. The top panel illustrates the same results found by \citetalias{JADES_DR3}, indicating that the standard correction adopted for the intrashutter source position in DR3 is insufficient. The bottom panel shows the wavelength calibration bias calculated in the spectra of DR4, where we used a new data-based wavelength correction as defined in equation~\ref{eq.wlzpcorr}.}
\label{f.prismoffset}
\end{figure}

\citetalias{JADES_DR3} also reported a wavelength-dependent discrepancy in redshift estimates within individual prism spectra: redshifts derived from emission lines at short wavelengths ($<2~\mu$m) were found to differ from those based on redward lines ($>3~\mu$m). Notably, this redshift discrepancy appears to correlate with the intra-shutter position of the target along the dispersion direction.
To further investigate this effect, we performed a similar quality assessment using a sample of galaxies in the redshift range $ 1 < z < 3 $, selected to have simultaneous detections of both \OIII and \HeI lines in the prism spectrum, each with a signal-to-noise ratio exceeding 10$\sigma$. Redshifts were estimated from the \HeI line, and the spectral offsets were defined as
\begin{equation}\label{eq.prismcal}
\Delta v = c \left(\frac{\lambda_{\rm [OIII]}}{ \lambda^{\rm rest}_{\rm [OIII]}(z_{\rm HeI}+1)}- 1\right),
\end{equation}

where $c$ is the light speed,  $z_{\rm HeI}$ is the redshift calculated from the helium line, and $\lambda_{\rm [OIII]}$ and $\lambda^{\rm rest}_{\rm [OIII]}$ are the measured and rest-frame line centroid of the oxygen line, respectively. Figure~\ref{f.prismoffset} shows the spectral offset, in units of velocity, as a function of the intra-shutter position, both before and after applying our empirical correction. Prior to the correction, we observed the same trend reported by \citetalias{JADES_DR3}. The new correction effectively removes the systematic dependence on intra-shutter location, significantly reducing the spectral offset between redshift measurements obtained at shorter wavelengths and those estimated at longer wavelengths. However, an offset of approximately 1000~km~s$^{-1}$, which is below the prism spectral resolution, remains visible in some individual galaxies.

\subsection{Accuracy on the absolute flux calibration}\label{s.flux_cal}

In this data release, we include all computed path-loss corrections by the pipeline in the final 1D and 2D spectra (see \S~\ref{s.limitations}). Using the intra-shutter position of the target and assuming a point-like source, the pipeline estimates two components of the pathloss correction: the ``geometrical'' loss, which accounts for light that does not pass through the aperture, and the ``diffraction'' loss, which arises from light diffracted by the aperture that is subsequently lost at the pupil plane of the instrument. 

As these corrections are based on the assumption that the target is not spatially extended, we performed a quality assessment by comparing the NIRSpec PRISM fluxes with those measured from NIRCam images at the same wavelength. We therefore used the latest version of the JADES NIRCam catalogue\footnote{https://archive.stsci.edu/hlsp/jades} (JADES collaboration in prep.) and cross-matched the catalogues of the targets using the coordinates associated with the 1D spectra with a searching radius of 0.1 arcsec. We then computed synthetic NIRCam photometry for all spectra at six different wavelengths using the nominal throughputs of F090W, F150W, F200W, F277W, F356W, and F444W filters, and compared it with the CIRC2 NIRCam photometries. The latter were estimated from a circular aperture of radius 0.15 arcsec, which is comparable to the open area of the NIRSpec shutters.

\begin{figure}
  \includegraphics[width=\columnwidth]{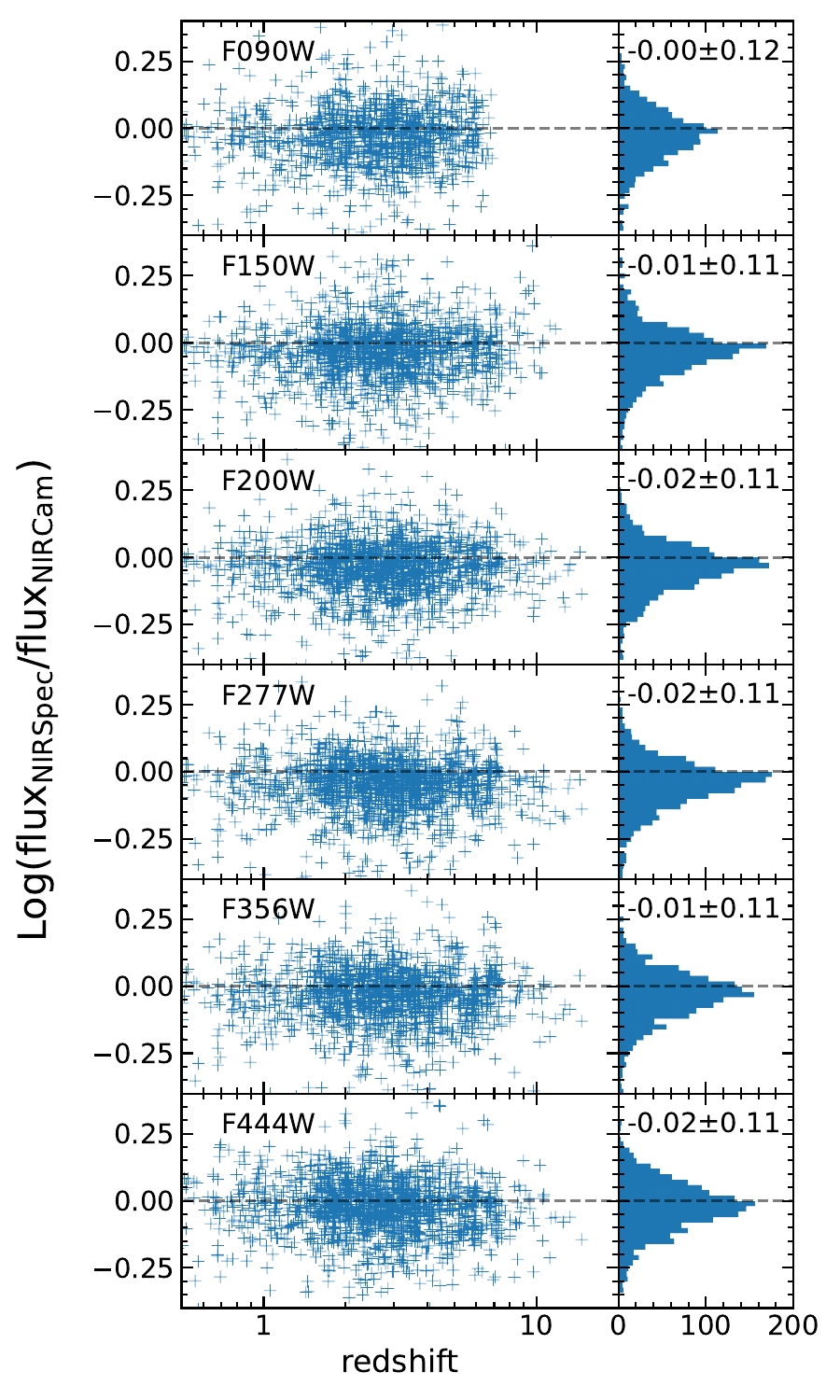}
  \caption{Left: ratio between the synthetic NIRCam photometry of NIRSpec spectra and CIRC2 NIRCam photometries from JADES NIRcam catalogue as a function of redshift for six different filters as reported in the plots. CIRC2 NIRCam photometries are extracted from a circular aperture of radius 0.15 arcsec. Right: distribution of the ratio in a logarithmic scale. The text reports the median and standard deviation of each distribution of values.}
  \label{f.fluxcal}
\end{figure}

Figure~\ref{f.fluxcal} shows the ratio of the synthetic NIRCam photometry of NIRSpec spectra and CIRC2 NIRCam photometry as a function of redshift for the six selected filters. We note that the flux level of the NIRspec spectra is consistent with the uncertainties of the NIRCam photometry. However, the intrinsic scatter of the distribution is about 0.11 dex at all wavelengths. This corresponds to a 25\% absolute flux accuracy of the NIRspec spectra. 

\subsection{Flux calibration between PRISM and Gratings}\label{s.flux_cal_inter}

\begin{figure}
  \includegraphics[width=\columnwidth]{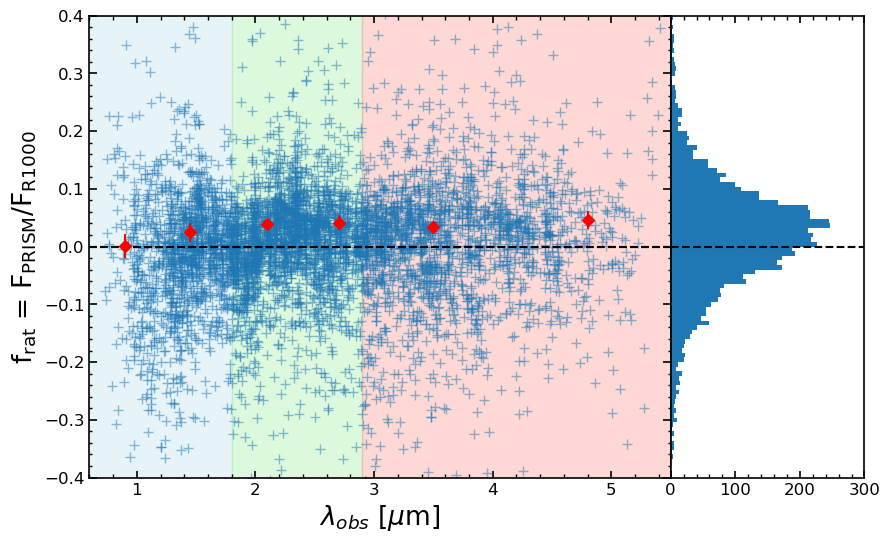}
  \caption{ Ratio of the fluxes of emission lines measured from the Gratings and PRISM as a function of the wavelength. We compare isolated emission lines: \OIIall, \OIIIall, (\Halpha+\NIIall), \Hbeta, [SIII] and [SIII]. The blue points show the individual measurements, while the red points show the mean in distinct bins. The blue, green, and red shaded regions indicate the three gratings G140m, G235m and G395m, respectively. We find a mean offset between the PRISM and Grating of 8\%. 
  }\label{f.flux.fcomp}
\end{figure}

In this section, we perform a detailed comparison of the flux calibration between the PRISM and medium gratings, as well as our comparison to the DR3. As the continuum is rarely detected in the grating spectra, we compare the measured emission line fluxes between the PRISM and grating observations, rather than to the NIRCam photometry as done above. For the emission line flux comparison, we chose \OIIIall, \Halpha+\NIIall, \Hbeta, [SIII]$\lambda$9062 and [SIII]$\lambda$9532 emission lines, as they are well isolated and bright for accurate comparison of the data releases and dispersers. We highlight the wavelength range of each of the three grating filters as shaded regions in Fig. \ref{f.flux.fcomp}. 

We note that in PRISM observations, there is an increased discrepancy in \OIIall between grating and PRISM due to the low resolution of the PRISM observations, and the effect that prominent continuum features near the Balmer limit can display a break \citetext{stellar Balmer break, e.g. \citealp{looser+2024}; or nebular Balmer jump, e.g. \citealp{cameron+2023c}}. When the resolution is insufficient, the strength of the break is degenerate with the flux of the nearest lines, with \OIIall being the most affected line. This is aggravated by using an incorrect value of the spectral resolution due to the object being smaller than the slit, which increases the resolution of the observations. All these issues can significantly affect the recovered \OIIall flux, particularly in the low-resolution regimes found at lower wavelengths \citep[e.g.][]{degraaff+2025}. 

Overall, we find a mean ratio of the PRISM and grating fluxes of $1.080\pm0.004$ with a standard deviation of 0.36, consistent with the findings of \citetalias{bunker+2023b} and \citetalias{JADES_DR3}. The flux ratio is constant across the wavelength range of the instrument with a value above 1.5$\mu$m of $\sim$1.09, showing that the individual gratings have a consistent flux calibration. Indeed, we see that the spectral regions where the grating spectra overlap the flux value and emission line are consistent. 
Given the excellent consistency of the PRISM with the NIRCam imaging, we consider the PRISM observations to be correct, with the grating observations being higher by $\sim$10\%. Although this difference is not enough to influence the majority of science cases, we note that combining PRISM and R1000 observations for science cases sensitive to line ratios (such as Balmer decrement and hence attenuation curves) needs to take into account the systematic uncertainties between the dispersers.

\begin{figure}
  \includegraphics[width=\columnwidth]{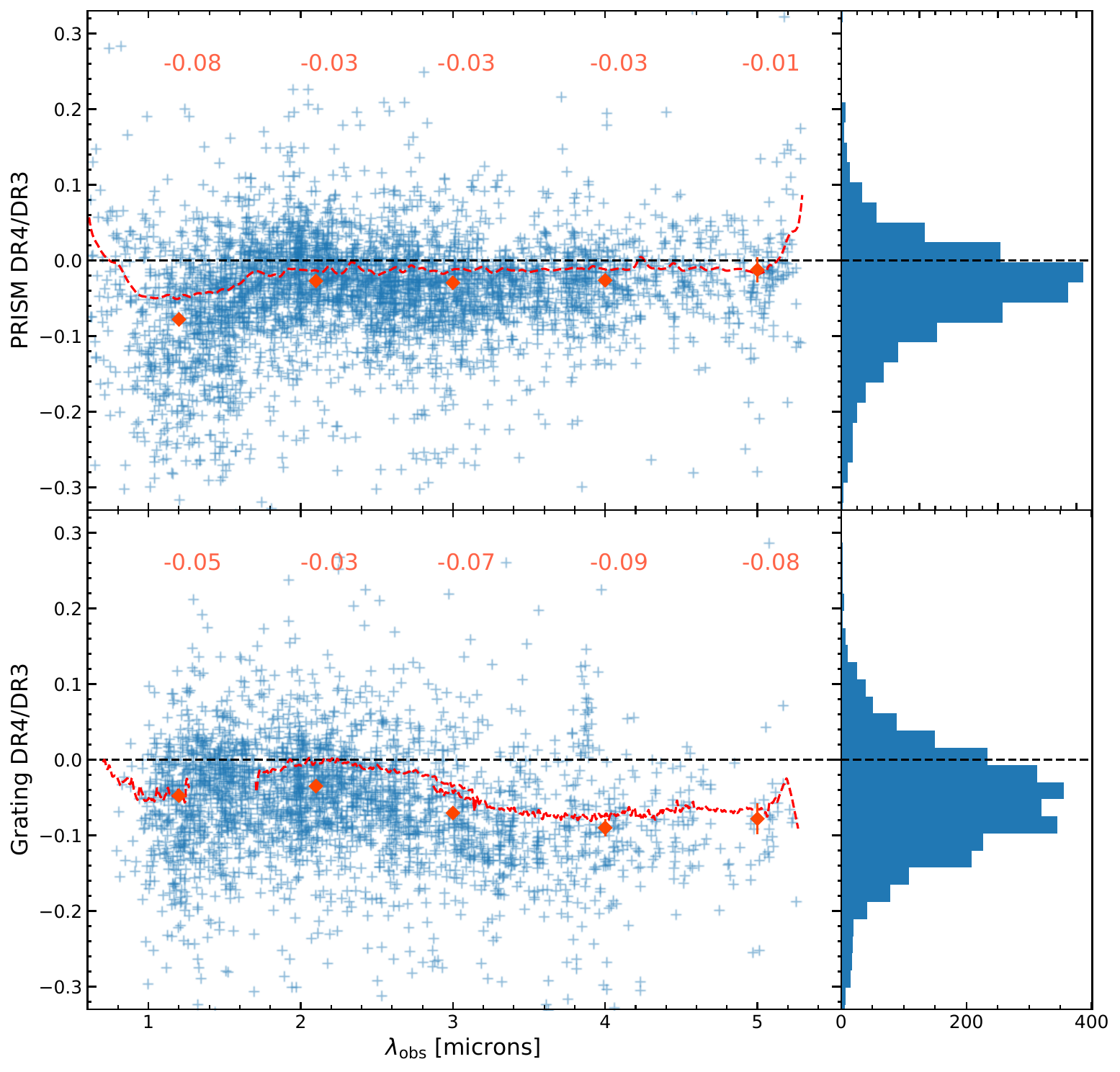}
  \caption{ Ratio of the fluxes from the DR4 and DR3 for PRISM fluxes (left panel) and Grating (right panel). The black points show the individual measurements. The red points show average ratios, which we also show as red numbers above the points. The dashed red line is showing the ratio of the SFlats and FFlats used in the data release, while the red points show the mean in distinct bins. The changes between DR4 and DR3 can be explained by the combination of new calibration files and new path loss corrections.}\label{f.flux.DR_comp}
\end{figure}

Given that we have re-reduced the entire spectroscopic survey with new context files (see \S~\ref{s.datared}), we need to investigate any changes between the previous data release (DR3) and the one presented in this work. In Figure \ref{f.flux.DR_comp} we compare the prism (left panel) and grating (right panel) fluxes released as part of DR3 and DR4 for the same emission lines as for Figure \ref{f.flux.fcomp}. We show the ratios between the new and the old calibration files as a red dashed line. We see that the difference in the data releases is purely driven by the change of the calibration files rather than any other changes. We note that there are a few objects with emission line ratios between DR3 and DR4 over a factor of 1.5 (75 emission lines in total). We note that the final ratio between the fluxes between DR3 and DR4 is a combination of both \texttt{Sflats} and \texttt{Fflats}, as well as new changes to pathloss corrections. Therefore, the new fluxes are not a simple multiplicative factor of the new calibration files but also depend on the source's location in the shutter, creating the wide scatter in the comparison and some large outliers. 

\subsection{Redshift combination and comparison: prism vs medium gratings}\label{s.redshift_quality}

\begin{figure}
  \includegraphics[width=\columnwidth]{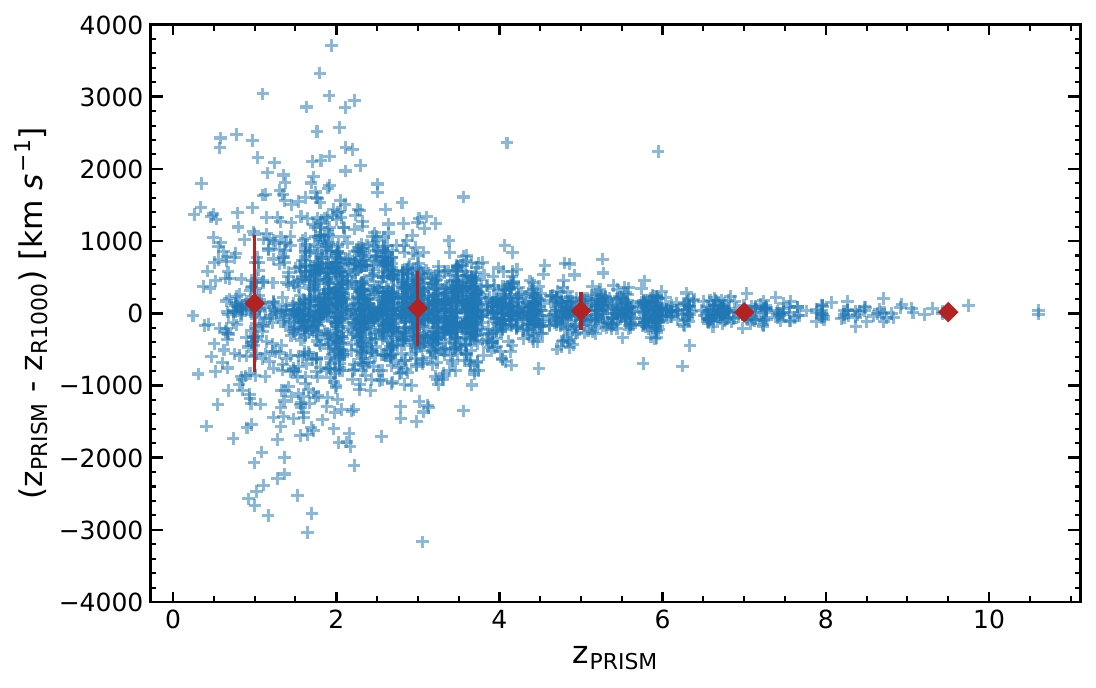}
  \caption{Comparison of redshifts between the prism and the medium-resolution gratings. We find a redshift-independent offset 72$\pm$8\kms. The red diamonds are the moving mean, with the errorbars showing the scatter in velocity offset between PRISM and R1000. The increased dispersion at low redshifts is expected from the strong dependence of the prism spectral resolution with wavelength.
  }\label{f.redshift.zcomp}
\end{figure}

For the final redshift estimate for an object, we combined the redshift measurement from both PRISM and Gratings to allow the most accurate version of the redshift. Whenever we have a strong emission line detection (i.e. 5~\textsigma) in the gratings, we adopt the redshift of this emission line and give it a redshift flag A. Using a single emission line is warranted because the Grating fits are done only for galaxies with a visually inspected redshift (see \citetalias{JADES_DR3}). We have manually verified that none of our objects have discrepant redshifts between PRISM and Gratings by more than $\Delta\,z = 0.05$, which would indicate misidentified emission lines. Furthermore, we note that any large offsets ($|\Delta\,z-\langle \Delta\,z\rangle|>0.015$ (Figure~\ref{f.redshift.zcomp}) were visually inspected, and are mostly due to uncertainties in the \Hbeta-\OIIIall blend and to low signal-to-noise data.

We assigned the redshift quality Flag B to any target for which we only have a PRISM detection of at least two independently fitted emission lines. For a combination of a strong continuum break and an emission line (i.e. less secure or precise redshifts), we assigned redshift flag C. An even lower class is reserved for redshifts identified as tentative in the visual inspection; in this case, we report the visual-inspection redshift (flag D). All other redshifts are assigned -1 (flag E). We summarise the redshift flags below: 

\begin{itemize}
  \item \texttt{[A]} Redshift from at least one emission line in the medium-resolution grating.
  \item \texttt{[B]} Redshift from two or more prism emission lines.
  \item \texttt{[C]} Redshift from the continuum, or from the continuum and a single prism emission line.
  \item \texttt{[D]} Tentative, from visual inspection.
  \item \texttt{[E]} No redshift.
\end{itemize}

\begin{figure}
  \includegraphics[width=\columnwidth]{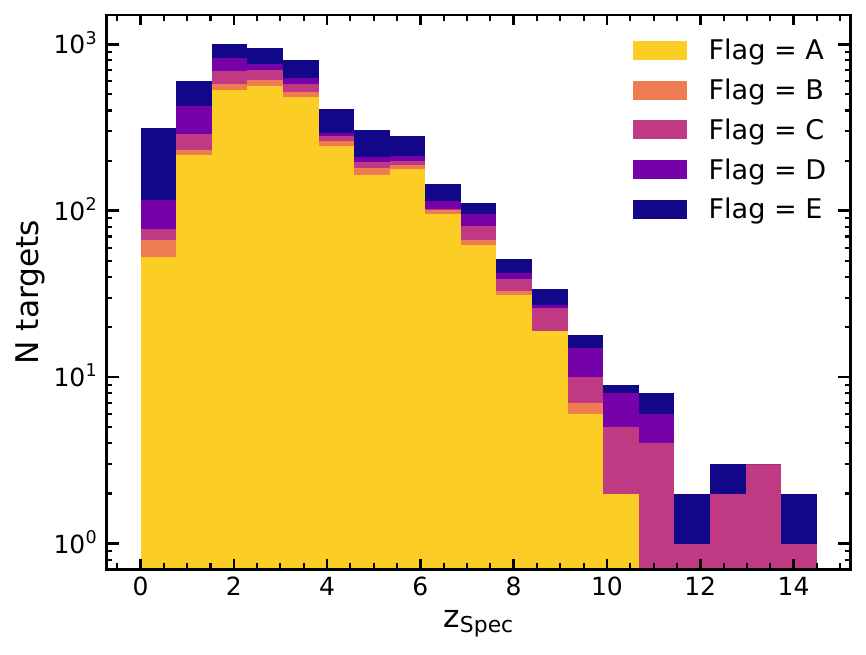}
  \caption{Redshift distribution of the sample, colour-coded by the final redshift flag (Section~\ref{s.quality}).
  We note a drop in redshift distribution at $z_\mathrm{Spec} \sim 9.75$; this reflects a similar drop in the
  distribution of photometric redshifts of the targets selected for observation. For flag "E" we use the photometric redshift for illustration.
  }\label{f.finalz}
\end{figure}

Note that the first three flags are the same as in \citetalias{bunker+2023b} and \citetalias{JADES_DR3}.  These measured redshifts and associated quality flags are also presented and used in \citetalias{DR4_paper1} to assess the success of our target selection strategy. We note that a full assessment of the redshift quality assessment and success of our strategy are presented in \citetalias{DR4_paper1}. The combined redshift distribution of the sample is shown in Figure~\ref{f.finalz}, colour-coded by flag. There is an accelerated decline in secure spectroscopic $z_\mathrm{Spec} \sim9.75$, consistent with the similar feature in the photometric redshift distribution of our source catalogue as mentioned by \citetalias{JADES_DR3}. 

The overall distribution of the spectroscopically measured redshifts and the redshifts flags are displayed in Figure~\ref{f.finalz} (we show only targets with flag A--C). To view the true evolution in the UV luminosity density from the spectroscopic sample, one must take account of the survey selection function.  The fraction of the underlying population sampled, as well as the redshift completeness and outlier fraction, are required for this level of calculation and are provided on a tier-by-tier basis in \citetalias{DR4_paper1}. 

\begin{figure}
  \includegraphics[width=\columnwidth]{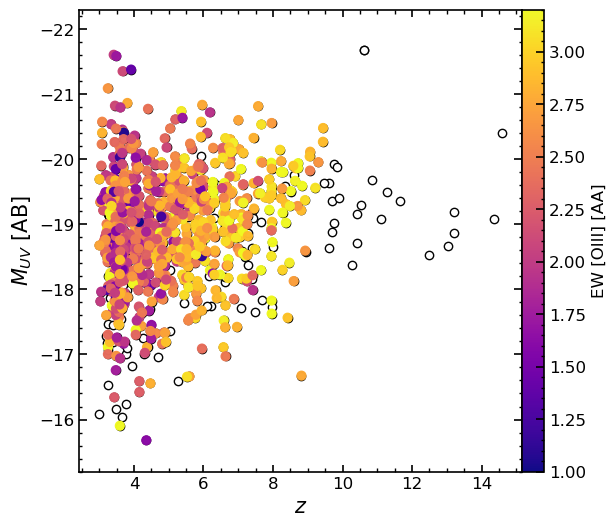}
  \caption{Redshift vs UV magnitude distribution of the sample, colour-coded by the equivalent width of \OIIIL (empty symbols are galaxies with no detected \OIIIL). Magnitudes were calculated directly from the prism spectra, using aperture corrections estimated by comparing the
  prism magnitude to the 0.35-arcsec radius magnitude (\texttt{CIRC5} in the catalogues) in the NIRCam filter nearest to rest-frame 1,500~\AA. \OIIIL falls outside of the NIRSpec wavelength range at $z\gtrsim10.0$. For more information on the EW of \OIIIall see \citet{boyett+2024}.
  }\label{f.zmuv}
\end{figure}

In Figure~\ref{f.zmuv} we show $M_\mathrm{UV}$ vs redshift, for the sample where magnitudes could be measured directly from the NIRSpec data; to this end, we used a nominal top-hat filter between rest-frame 1,400 and 1,600~\AA, following the procedure outlined in \citet{Saxena+2024}. We colour-coded our points with the equivalent width of \OIIIL, measured directly on the prism data (empty symbols are galaxies with no detected \OIIIL). We do not see any evidence for a trend between the EW of the \OIII and the $M_\mathrm{UV}$ at fixed redshift. 

In addition, we see a lack of galaxies fainter than 29 mag at redshifts lower than $z_\mathrm{Spec}\lesssim 2$ and higher than $z_\mathrm{Spec}\gtrsim 9$; which, in addition to the survey selection function, is driven by sensitivity as follows. At low redshifts, NIRCam photometry becomes less able to clearly distinguish line excesses, because the spacing of strong emission lines reduces as $1+z$; the lower sensitivity of NIRSpec at wavelengths $\lambda < 1~\mum$ compounds the problem. At redshifts higher than $z_\mathrm{Spec}\approx 9.5$, instead, the strongest emission lines (\OIIIall) are redshifted out of the NIRSpec coverage, so redshift measurements rest solely on the \Lyalpha drop and on less prominent lines -- both of which are harder to detect in faint targets.

In Figure~\ref{f.redshift.zcomp} we compare the redshift measurements from the PRISM and from the medium gratings, for emission lines with SNR$>$10. In order to assess the quality of our redshifts we defined $\Delta v/c \equiv (z_\mathrm{prism} - z_\mathrm{gratings}$)/(1+$z_\mathrm{prism}$), and we find a mean offset of $72\pm 8$ km/s, significantly less than the values derived by \citetalias{bunker+2023b} and \citetalias{JADES_DR3}. We note that the scatter of this offset has a strong dependency on wavelength, or more importantly, on the native pixel size of the NIRSpec instrument, as we show with the errorbars in Figure~\ref{f.redshift.zcomp}.


\section{Using the NIRSpec data products}\label{s.limitations}

\subsection{Data products and Tables}\label{s.products}

In this section, we provide details of the data products made public in this data release, their use and their limitations. The spectra are published in the same format as described in \S~\ref{s.hlsp} 
on JADES website\footnote{\url{https://jades.herts.ac.uk/DR4} \& \url{https://jades-survey.github.io/scientists/data.html}}. The catalogues prepared for this data release are published 
on the JADES website and through the JADES online database (see \S~\ref{s.database})\footnote{\url{https://jades.herts.ac.uk/search/}}. The objects are in the same order in the tables/hdu extensions and are organised as follows: 

\begin{enumerate}
  \item \textbf{\texttt{Obs\_info}}: The master catalogue contains the basic properties for each of the targets in our observations. We summarise each of the columns in Table \ref{tab:Master}. 
  \item \textbf{PRISM Flux tables (3 pixel \& 5 pixel extractions)}: This is the table of measured fluxes in 3 and 5 pixel extractions of the PRISM spectra measured using methods described in \S~\ref{t.prism}. The structure of the columns is described in Table~\ref{t.prism}.  
  \item \textbf{R1000 Flux tables (3 pixel \& 5 pixel extractions)}: Table of the measured fluxes from the medium gratings. We describe the method of measuring the fluxes in \S~\ref{s.r1000} and we describe the table structure in Table~\ref{grating.eml_structure}. 
\end{enumerate}

We note that for the upload to the HLSP website, we are required to split the multi-HDU FITS table into separate \texttt{.fits} tables and also split the observation into separate tables for the GOODS-S and GOODS-N fields. The original tables can be downloaded at the JADES website or use our online database.

\subsection{JADES online database}\label{s.database}

In order to make our JADES NIRSpec data release more accessible and allow fast filtering and visualisation of the data, we built an online graphical interface. Within the interface, the user can search by sky coordinates (with a user-defined search radius), redshift range, redshift quality flags (described in \S~\ref{s.redshift_quality}), absolute UV magnitude and emission line fluxes (H$\alpha$, H$\beta$ and \OIIall, \OIIIall). The user can then search and display the target selection along with available NIRCam images, R1000 fits of H$\alpha$ and H$\beta$, PRISM spectrum, link to the \texttt{FITS\_map} - NIRCam image visualisation and link to the HLSP portal to download the available data. This online table with the displayed columns can be downloaded as a \texttt{.csv} file, or the user can download the entire JADES multi-hdu fits table using the \texttt{Download table} button. Using this graphical interface, the users can search, filter and download the sample they are interested in.

\begin{figure*}
  \includegraphics[width=0.8\paperwidth]{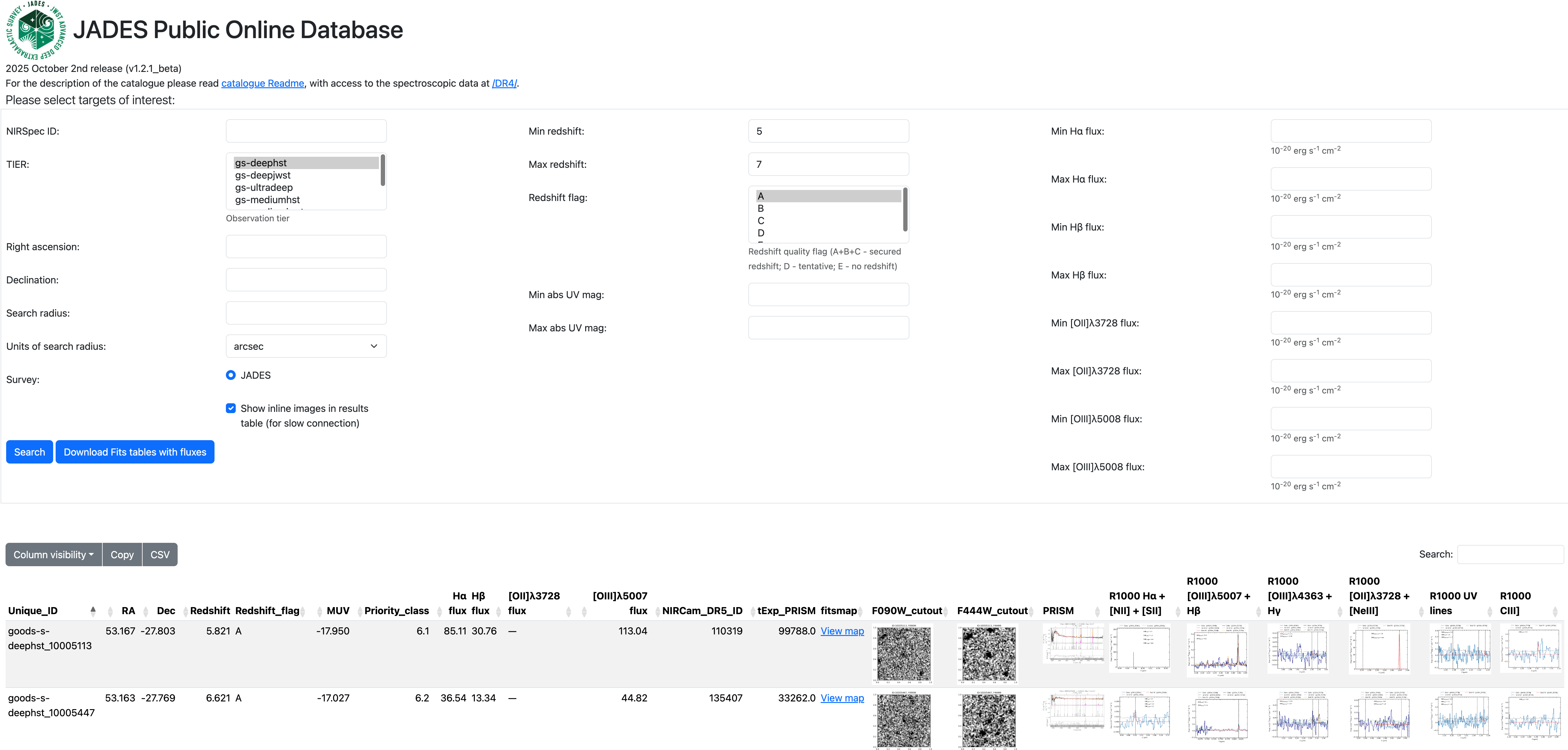}
  \caption{Image of the graphical interface of the JADES online database designed to visualise the JADES data.
  }\label{f.database}
\end{figure*}

\subsection{Limitations of the data release}

In this section, we provide a list of limitations of the current data release. We note that our data reduction and analysis are optimised for compact sources which are common at high redshift, the primary goal of JADES. The users of our data release are encouraged to consider these points carefully:

\begin{itemize}
  \item \textbf{Gold samples} - The JADES survey's selection process contains many different selection criteria and classes, resulting in  overall inhomogeneous sample, however, within the individual tiers and classes the selection was designed based on UV apparent magnitude or NIRCam F444W filter flux. However, the original allocated classes were determined from the best photometry available at the time, and the underlying catalogues have since evolved (see \citetalias{DR4_paper1}). For this reason we defined two ``gold'' samples based on spectroscopic redshifts (classes A, B, and C), applying consistent rest-UV and F444W selections using the most up-to-date photometry covering distinct redshift ranges: $z>5.7$ for the UV-selected sample and $1.5<z<5.7$ for the F444W-selected sample (for full description see \citetalias{DR4_paper1})). With these gold samples it is possible to reconstruct the UV luminosity function and other completeness studies. We note that we deliver the UV and F444W gold samples based on DR3 NIRCam photometry (\citetalias{JADES_DR3}) and upcoming DR5 NIRCam photometry (Johnson and JADES collaboration et al. in prep). 

  \item \textbf{Aperture correction.} Our aperture corrections are optimised for compact sources and assume that the targets have a point source morphology, i.e. the spectra and the colours are inaccurate for extended sources. When estimating emission line ratios over a large wavelength range (such as  \Halpha/\Hbeta, \OIIIL/\OIIall, \Paalpha/\Halpha), we recommend using aperture corrections derived from the photometry, using the relevant NIRCam images included in the published data files for this purpose. Targets that are extended over more than one shutter should be considered with particular care, or even excluded (see Background subtraction).

  \item \textbf{3 and 5 pixel extraction:} We extract the 1D spectra from the 2D spectra using two separate extractions - 3 and 5 pixels. We note that for compact, barely resolved sources, the 3 pixel extraction is more optimised for unresolved and close to unresolved sources, and the user might get better SNR compared to the 5 pixel extraction. We encourage the users to compare the measured fluxes in 3 and 5 pixel extractions to verify that they account for the total flux of the emission line in the target. 

  \item \textbf{Background subtraction:} Similarly to the aperture corrections, our background subtraction is optimised for compact sources; while shutters affected by contaminants are pre-identified and not considered in the subtraction, shutters affected by the same source cause self-subtraction. For these sources, the users are encouraged to use the 2-nod extraction spectra (see description of data products).

  \item \textbf{Noise spectrum:} Our uncertainties on the flux values are based on variance-conserving resampling, to minimise the effects of correlated noise \citep{dorner+2012} and hence are considered conservative \citep[e.g.][]{witstok+2025}. A full analysis of correlated noise in NIRSpec will be presented in a future work (P.~Jakobsen, in~prep.)
  
  \item \textbf{Wavelength calibration:} The updated wavelength calibration reduces the spectral offset between prism and grating data reported in DR3, leading to improved alignment in DR4. The calibration uncertainties are on the order of one-tenth of the spectral resolution of the disperser used in the observations.  Larger uncertainties occur for spatially extended galaxies or when the intra-shutter position of the source is poorly constrained. 

  \item  \textbf{Spectral overlap:} The PRISM observations were set up so that we avoid any spectral overlaps. However, for grating observations, there can be some overlap in the spectra for low-priority class observations. The users should take care to ensure these are not misinterpreted.
  
  \item \textbf{Flux calibration.} We determine that the absolute calibration of the PRISM observations is accurate within $\sim$25 \%. The relative flux calibration between the prism and gratings is accurate to within 10~per cent, and depends on the wavelength (e.g., Figs.~\ref{f.fluxcal}, \ref{f.flux.fcomp}, \ref{f.flux.DR_comp}). The user is encouraged to consider this problem when measuring flux ratios. We particularly note that estimating Balmer ratios across different gratings or between PRISM and gratings can lead to inaccurate estimates of dust extinction. In addition, the wavelength-dependent flux discrepancy between PRISM and gratings suggests that emission-line fluxes within the same disperser may also suffer from flux calibration issues.

\end{itemize}

\section{Conclusions}\label{s.conclusions}

In this work, we present the final JADES NIRSpec Data release which includes data obtained up to January 2025 in the two GOODS fields. The spectra include medium-depth, deep-depth and ultra deep-depth observations ($\sim$2--$\sim$70 hours on source) up to redshift $z \sim 14$, reaching the deepest unlensed spectroscopic observations to date, with up to 70~hours on source. The sample size of 5,190 galaxies and the data quality of spectra are a testament to the success of the MSA mode of the \jwst/NIRSpec telescope and instrument. 

This paper accompanies the release of the final calibrated spectra and the accompanying NIRCam image cutouts for each target, along with catalogues containing spectroscopic redshifts, emission line fluxes, and observational information. The large sample has enabled us to investigate remaining calibration challenges of the NIRSpec instrument within the data reduction: i) mismatch in the wavelength calibration between PRISM and gratings; ii) flux calibration between the used dispersers; and iii) background subtraction and slit-loss corrections appropriate for unresolved and extended sources. To make advancements in this section of the calibration, future calibration programmes are necessary to remove these last remaining challenges from the data. The spectra, photometry and survey completeness have been used to recover the UV luminosity function in the companion \citetalias{DR4_paper1}.

Future programmes aiming to significantly increase the sample size of the general galaxy population at similar redshift and stellar mass range to JADES will require a significant investment of \jwst time, or potentially a novel observing strategy with higher observing efficiency. However, regular GO programmes should be able to observe large samples of specific rare types of objects, such as (e.g., high-redshift quiescent galaxies, extremely reddened galaxies, little red dots, $z>8$ \Lyalpha emitters), where the  JADES survey has only scratched the surface. 

In the meantime, the JADES survey represents the largest sample of galaxies at high redshift covering the full 0.6--5.5~\mum with both PRISM and medium-resolution grating spectroscopy. Combining this spectroscopic data with deep medium and wide-band imaging enables the broader astronomical community to study galaxy evolution from the Cosmic Dawn to the Cosmic Noon. 

\section*{Acknowledgements}
This work is based on observations made with the NASA/ESA/CSA James Webb Space Telescope. The data were obtained from the Mikulski Archive for Space Telescopes at the Space Telescope Science Institute, which is operated by the Association of Universities for Research in Astronomy, Inc., under NASA contract NAS5-03127 for JWST. These observations are associated with programmes 1210, 1180, 1181, 1286, 1287 and 3215.
JS, RM, FDE, and GCJ acknowledge support by the Science and Technology Facilities Council (STFC), ERC Advanced Grant 695671 ``QUENCH" and the UKRI Frontier Research grant RISEandFALL. RM also acknowledges funding from a research professorship from the Royal Society.
S.C. and E.P. acknowledge support from the European Union (ERC, WINGS,101040227)
ECL acknowledges support of an STFC Webb Fellowship (ST/W001438/1).
The Cosmic Dawn Center (DAWN) is funded by the Danish National Research Foundation under grant DNRF140.
AJB, AS and AJC acknowledge funding from the "FirstGalaxies" Advanced Grant from the European Research Council (ERC) under the European Union’s Horizon 2020 research and innovation programme (Grant agreement No. 789056).
WMB gratefully acknowledges support from DARK via the DARK fellowship. This work was supported by a research grant (VIL54489) from VILLUM FONDEN.
DJE is supported as a Simons Investigator and by JWST/NIRCam contract to the University of Arizona, NAS5-02115.  Support for program \#3215 was provided by NASA through a grant from the Space Telescope Science Institute, which is operated by the Association of Universities for Research in Astronomy, Inc., under NASA contract NAS 5-03127. 
ZJ, MR and BDJ, CNAW are supported by JWST/NIRCam contract to the University of Arizona, NAS5-02115.
MP acknowledges support through the grants PID2021-127718NB-I00 and RYC2023-044853-I, funded by the Spain Ministry of Science and Innovation/State Agency of Research MCIN/AEI/10.13039/501100011033 and El Fondo Social Europeo Plus FSE+.
PGP-G acknowledges support from grant PID2022-139567NB-I00 funded by Spanish Ministerio de Ciencia e Innovaci\'on MCIN/AEI/10.13039/501100011033, FEDER, UE.
BER acknowledges support from the NIRCam Science Team contract to the University of Arizona, NAS5-02115, and JWST Program 3215.
BRP acknowledges support from grant PID2024-158856NA-I00 funded by Spanish Ministerio de Ciencia e Innovación MCIN/AEI/10.13039/501100011033 and by “ERDF A way of making Europe”
MSS acknowledges support by the Science and Technology Facilities Council (STFC) grant ST/V506709/1.
H\"U acknowledges funding by the European Union (ERC APEX, 101164796). Views and opinions expressed are however those of the authors only and do not necessarily reflect those of the European Union or the European Research Council Executive Agency. Neither the European Union nor the granting authority can be held responsible for them.
The research of CCW is supported by NOIRLab, which is managed by the Association of Universities for Research in Astronomy (AURA) under a cooperative agreement with the National Science Foundation.
The authors acknowledge use of the lux supercomputer at UC Santa Cruz, funded by NSF MRI grant AST 1828315.
JWST/NIRCam contract to the University of Arizona NAS5-02115
JW gratefully acknowledges support from the Cosmic Dawn Center through the DAWN Fellowship. The Cosmic Dawn Center (DAWN) is funded by the Danish National Research Foundation under grant No. 140.
The authors acknowledge use of the lux supercomputer at UC Santa Cruz, funded by NSF MRI grant AST 1828315.

\section*{Data Availability}

The datasets were derived from sources in the public domain: JWST/NIRSpec MSA and JWST/NIRCam data from MAST portal - \url{https://mast.stsci.edu/portal/Mashup/Clients/Mast/Portal.html} as well as our own reduction and analysis at \url{https://jades.herts.ac.uk/DR4/} and \url{https://jades.herts.ac.uk/search/}



\appendix
\section{Extension of flat field curve}\label{app_ext}

As presented in Section~\ref{s.datared}, the pipeline requires three flat field transmission curves, D-FLAT, S-FLAT, and F-FLAT, to process the MOS data and obtain the final flux-calibrated 1D and 2D spectra. The JWST CRDS provides the reference files for the flat field for each disperser/filter configurations. However, all flat field correction curves include in the reference files are defined only within the nominal wavelength ranges but the D-FLAT correction curve is already defined from 0.6$\mu$m to 5.85$\mu$m. Here, we discuss how we extended the curves of the F-FLAT and S-FLAT. 

\begin{figure}
\includegraphics[width=0.45\textwidth]{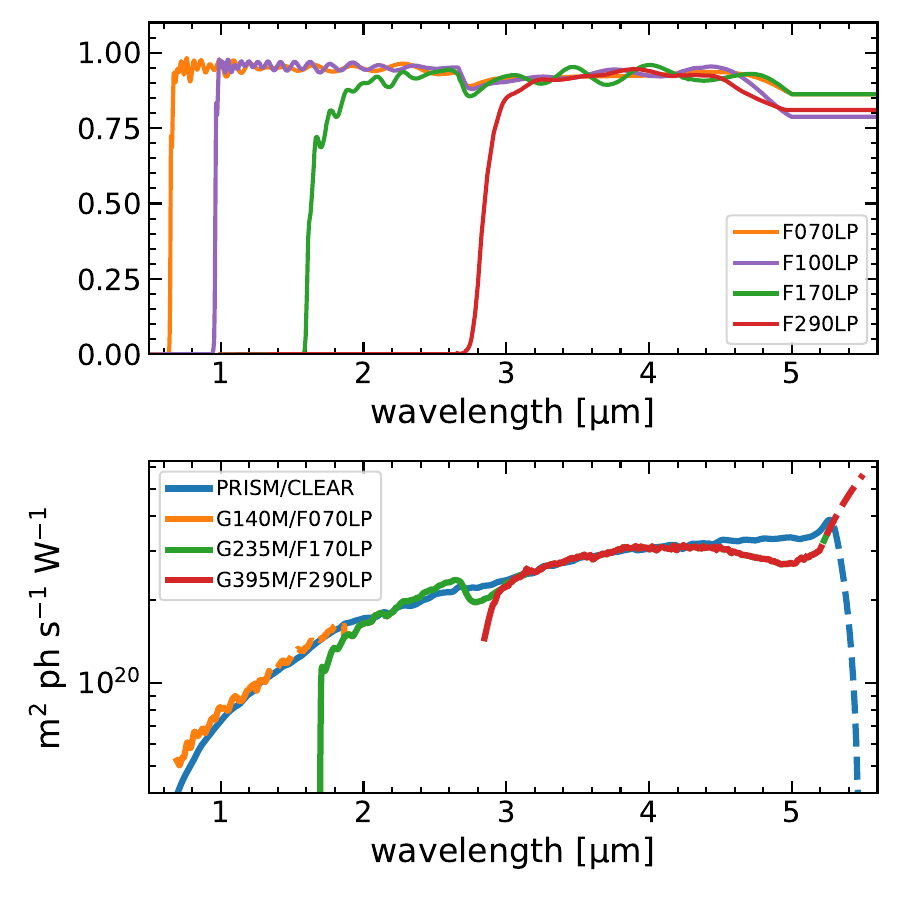}
\caption{The top panel shows the transmission throughputs of the long-pass filters used in NIRSpec observations with the gratings. The bottom panel displays the F-FLAT curves applied to calibrate the data in this release. Dashed lines indicate the extrapolated portions of each curve, which extend the profiles beyond their nominal wavelength ranges.}
\label{f.fflt}
\end{figure}

The F-FLAT curves depend mainly on the filters used in the selected disperse/filter configurations. R100 observations are obtained with the configuration PRISM/CLEAR that includes a bandpass filter, with both a cut-on and cut-off wavelength. The nominal maximal wavelength is 5.3$\mu$m after which the transmission of the bandpass filter drops rapidly. However, we noted that a faint signal at longer wavelength is recovered in both detectors. Therefore, we extended the wavelength range by performing a linear fitting of the curve between 5.2 and 5.3$\mu$m and extending the profile up to 5.5$\mu$m. The extended profile is shown in Figure~\ref{f.fflt} as a dashed line. R1000 observations use long-pass filters, which have only a cut-on wavelength. The transmission throughputs of these filters are similar across overlapping wavelength ranges (see top panel of Fig.~\ref{f.fflt}). Therefore, to extend the wavelength range of the reference files, we used the F-FLAT curves from the G140/F100LP and G395M/F290LP configurations as proxies for the G140/F070LP and G235M/F170LP configurations, respectively, beyond the nominal wavelength range. In the case of G395M/F290LP we performed a linear extrapolation of the curve up to 5.5 $\mu$m with the same method used for the PRISM/CLEAR. The final F-FLAT curves for the R1000 configurations are shown in Fig.~\ref{f.fflt}.

\begin{figure}
\includegraphics[width=0.5\textwidth]{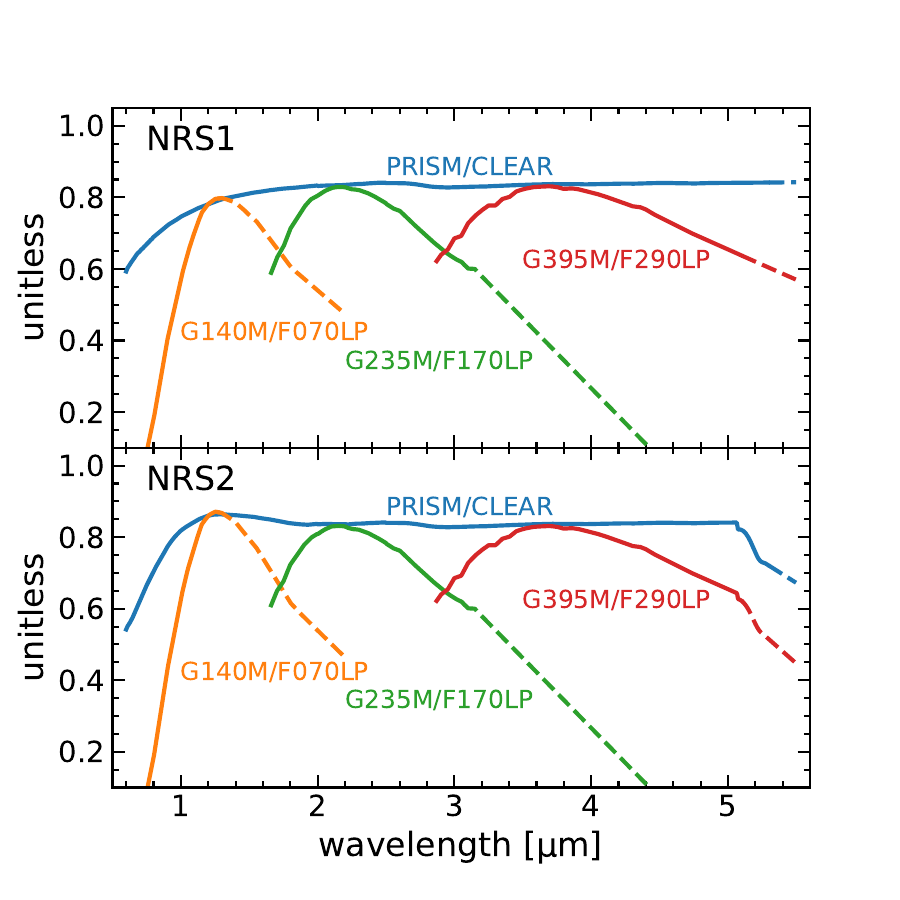}
\caption{S-FLAT curves for each configuration and detector (NRS1 and NRS2) adopted to process the spectra of this data release. Dashed lines indicate the extrapolated portions of each curve, which extend the profiles beyond their nominal wavelength ranges.}
\label{f.sflt}
\end{figure}

The S-FLAT curves represent the photon conversion efficiency as a function of wavelength and depend primarily on the properties of the disperser. Since the G140M/F070LP and G140M/F100LP configurations use the same grating, we used the S-FLAT curve from the latter to extend the reference file profile for the G140M/F070LP configuration. For the other three configurations—PRISM/CLEAR, G235M/F170LP, and G395M/F290LP—we fitted the S-FLAT curve over the final 0.1~$\mu$m of the nominal wavelength range using a first-order polynomial, and then used the best-fit model to extrapolate the curve up to 5.5~$\mu$m. The final curves are shown in Figure~\ref{f.sflt}. We stress that the extrapolated curves are just a first-order approximation because we expect that the real curve include ``bump'' and ``wiggles'' in the profile, as we see in the nominal wavelength range of each configuration.

\section{Data-derived path-losses correction}\label{app_pthl}

The JADES NIRSpec pipeline, like the standard STScI pipeline, includes a step dubbed as ``pathloss'', which calculates and applies corrections for each target in the MSA mask to account for signal losses along the optical path in the MOS spectroscopic mode. This correction is calculated assuming that all targets are point-like sources and using a PATHLOSS reference file that contains correction factors as functions of source position in the shutter aperture and wavelength. In the previous JADES data release (DR3;  \citetalias{JADES_DR3}), the pathloss correction was estimated using the PATHLOSS reference file \textit{jwst\_nirspec\_pathloss\_0005.fits}, which is based on detailed Fourier-optics simulations of the NIRSpec instrument carried out prior to the launch of JWST \citep{jakobsen+2022}. 

For this data release, we opted to compute new pathloss corrections based on observations from JWST commissioning program 1133 (PI: Catarina Alves de Oliveira), which targeted 20 isolated stars in the Large Magellanic Cloud using the MOS mode. These stars were selected to be evenly distributed across the MSA field of view. Each star was observed 20  times, with its position varying within the shutter in each exposure. All stars have at least one exposure where the star is centered in the shutter, where the pathloss correction is expected to be negligible. This setup enables the construction of a shutter slit-loss map by comparing the relative flux differences between spectra taken with the star at the shutter center and those at different positions. Therefore, we processed the data with the standard pipeline but without applying any pathloss correction and combining the final spectra. Here, we present how the data-based pathloss correction was derived from the individual 1D spectra of each star and exposure.

\begin{figure}
\includegraphics[trim={0 0cm 0 0},clip,width=0.5\textwidth]{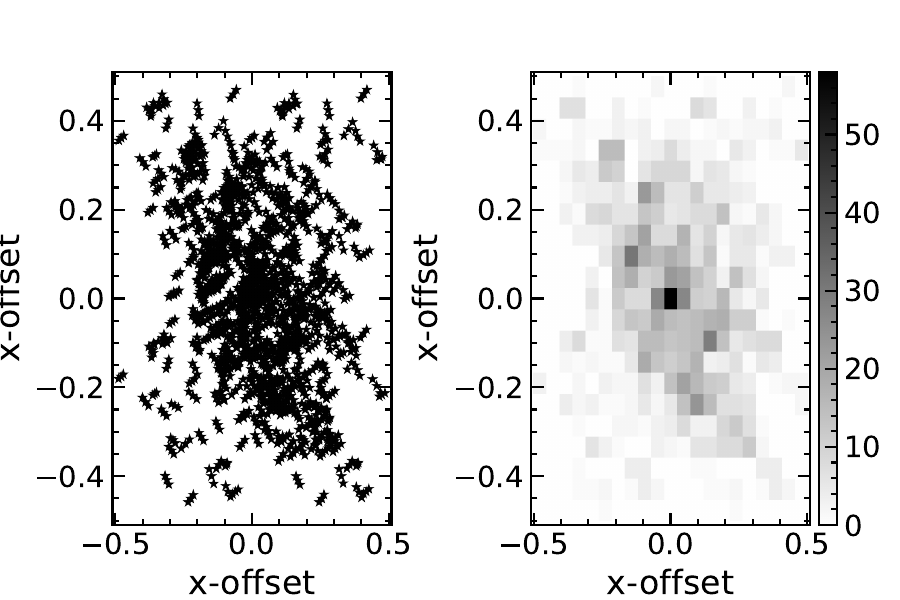}
\caption{Placement of stars of program 1133 into operable MSA shutters. Left panels show each star's position in its own shutter, all plotted together in a single virtual shutter. The right panel presents a 2D histogram illustrating the distribution of stellar positions within this virtual shutter. }\label{f.pthl1}
\end{figure}

Figure~\ref{f.pthl1} illustrates the intra-shutter positions of all stars and exposures from the commissioning program 1133. The bi-dimensional histogram in the right panel shows the number of observations per 2D bin across the virtual shutter. The distribution reveals that the coverage of the shutter field of view is not uniform. The central region within 0.15 of the shutter pitch is well sampled, with more than 20 observations per bin, while the outer regions are more sparsely populated and, in some cases, entirely uncovered.

To improve the sampling, we assumed that the pathlosses correction map is fully symmetric on both axes. This means that if the $f(x,y)$ is the correction at the intrashutter position $(x,y)$, our assumption implies $f(-x,-y)=f(-x,y)=f(x,-y)=f(x,y)$. This allowed us to improve the sampling by a factor of four.

\begin{figure}
\includegraphics[trim={0 1cm 0 0},clip,width=0.48\textwidth]{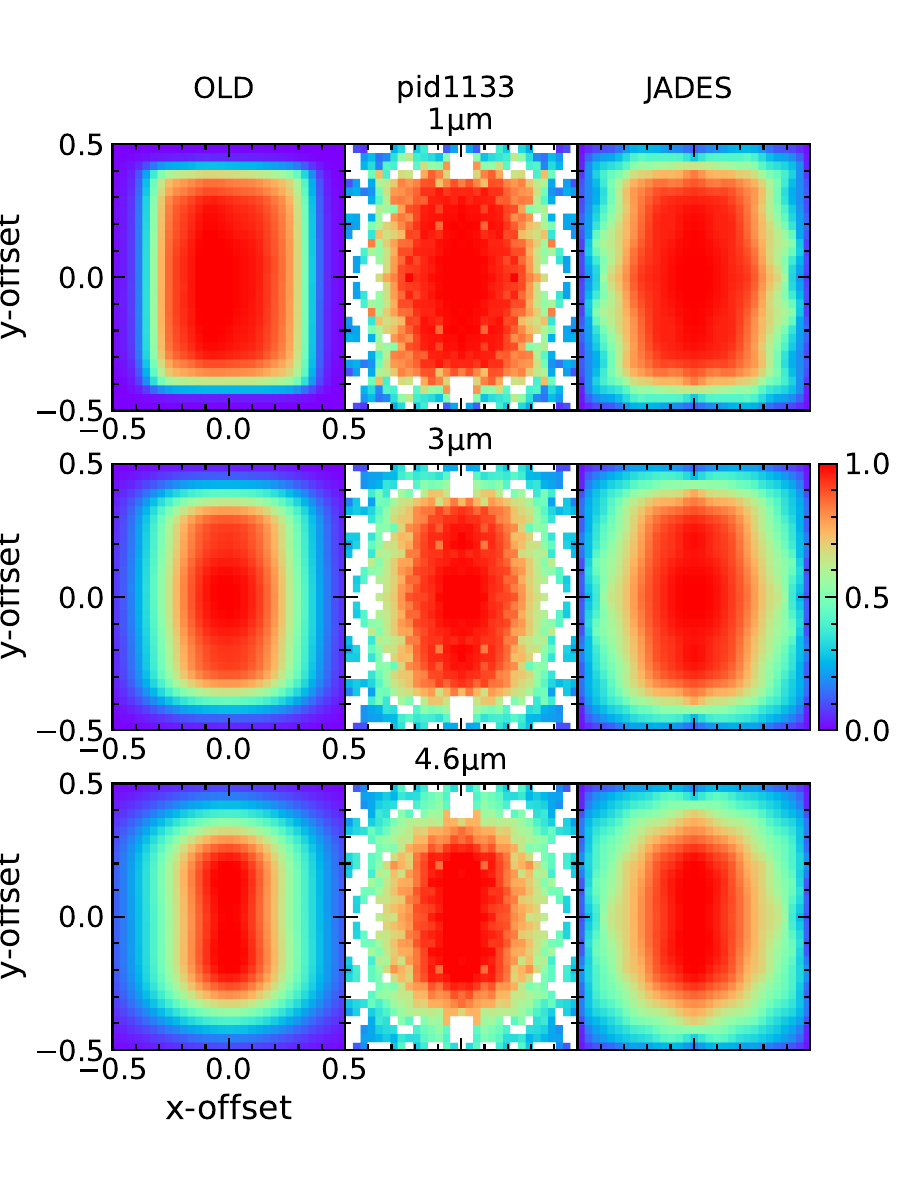}
\caption{Path loss correction maps for a point source at 1$\mu$m, 3$\mu$m, and 4.6 $\mu$m. The first column illustrates the correction maps adopted in DR3 and corresponding to the reference file \textit{jwst\_nirspec\_pathloss\_0005.fits}. The second column shows the estimates from the observations. The third column reports extrapolated from the observations after performing a smoothing and linear interpolation in the region with no measurements. These are the correction maps adopted in this data release.  }
\label{f.pthl_2}
\end{figure}

We created a reference file similar to that delivered by STScI which is a 3D datacube that samples the intrashutter position with a uniform 2D $31\times31$ grid and a uniform wavelength array from 0.6$\mu$m to 5.5$\mu$m of 20 elements is used for the third axis of the cube. For each intra shutter position beam we estimated the pathloses correction as the median of the ratios between spectra at the center of the shutter and the spectra obtained at the specific off-center position. The results are shown in the second column from the left of Figure~\ref{f.pthl_2} at 1$\mu$m, 3$\mu$m, and 4.6$\mu$m. The empirical maps are somewhat different from the optical Fourier prediction (first column of the figure), but they have similar behaviour.

Despite the assumption of symmetry along the x- and y-axes, the empirical maps still contained some bins with missing values. To address this, we performed a 2D linear interpolation to estimate the missing measurements, followed by a Gaussian smoothing with a standard deviation of 0.5 bins to suppress spurious fluctuations across the map. The final maps are shown in the third column of Figure~\ref{f.pthl_2}.

\begin{figure}
\includegraphics[trim={0 1.5cm 0 0},clip,width=0.5\textwidth]{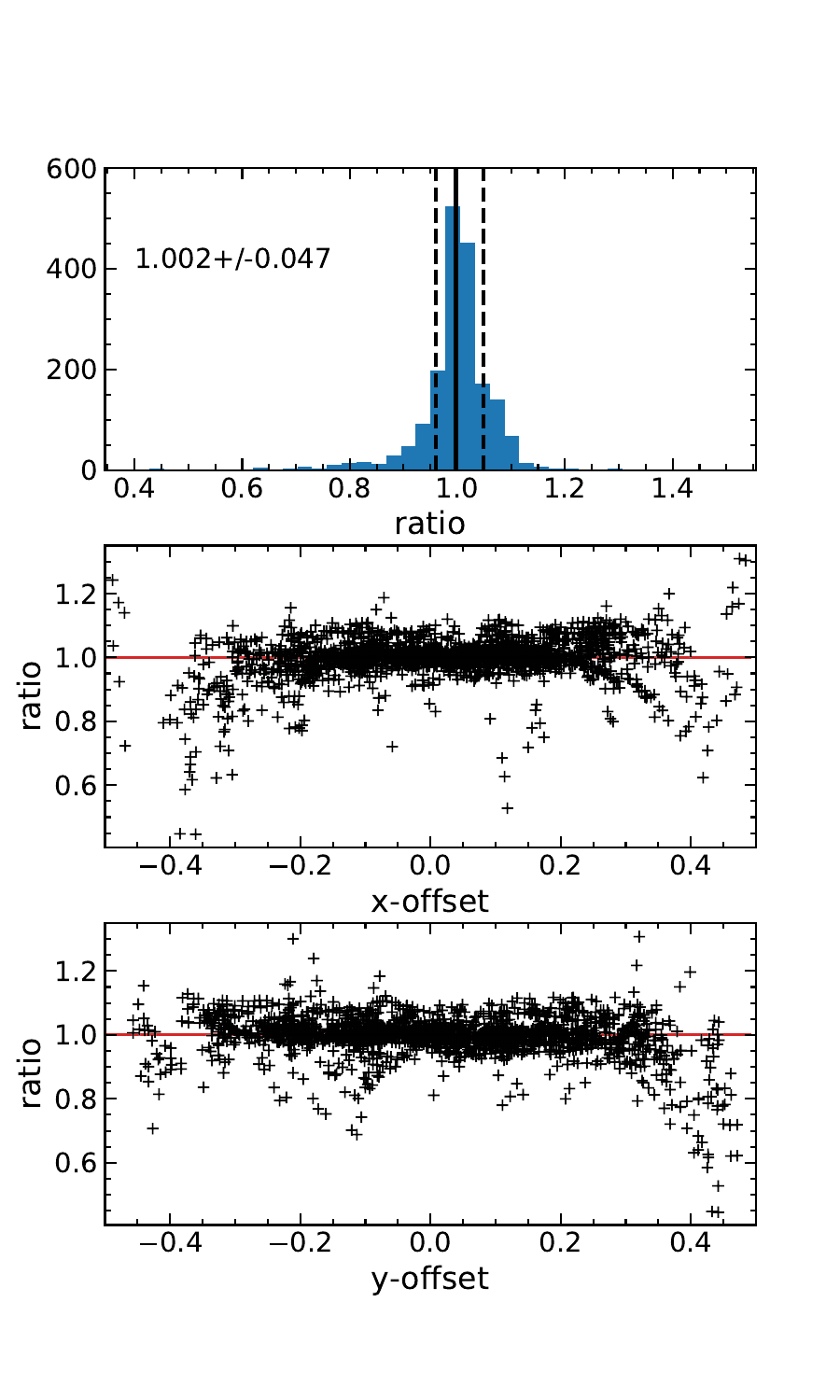}
\caption{Ratios at 3$\mu$m between spectra at the center of the shutter and the spectra of obtained at the specific off-center position. The top panel shows the distribution of the ratio estimates. The text reports the corresponding median and standard deviation values. The middle and bottom panels illustrate the ratio estimates as a function of the intra-shutter position along the x- and y-directions, respectively. }\label{f.pthl_3}
\end{figure}

We validated the effectiveness of our data-based pathloss correction maps by reprocessing the data from commissioning program 1133 using the updated pathloss reference file. To assess the accuracy of the corrections, we estimated the flux ratios between spectra obtained at the center of the shutter and those acquired at specific off-center positions. Figure~\ref{f.pthl_3} presents the distribution of ratio measurements at 3$\mu$m. On average, the new pathloss corrections are quite accurate, with uncertainties below 5\%. However, when analyzing the ratio estimates as a function of the intra-shutter position, we observe increased scatter toward the edges of the shutter (middle and bottom panels of Figure~\ref{f.pthl_3}). In particular, the corrections become less reliable when the target is located at a relative distance greater than 0.35 times the shutter pitch along either axis. This reduced accuracy is primarily due to the limited number of calibration observations at off-center positions near the shutter edges.

\section{Wavelength offset of  MAST final products}\label{app_mast_wl}

\begin{figure}
\includegraphics[width=0.45\textwidth]{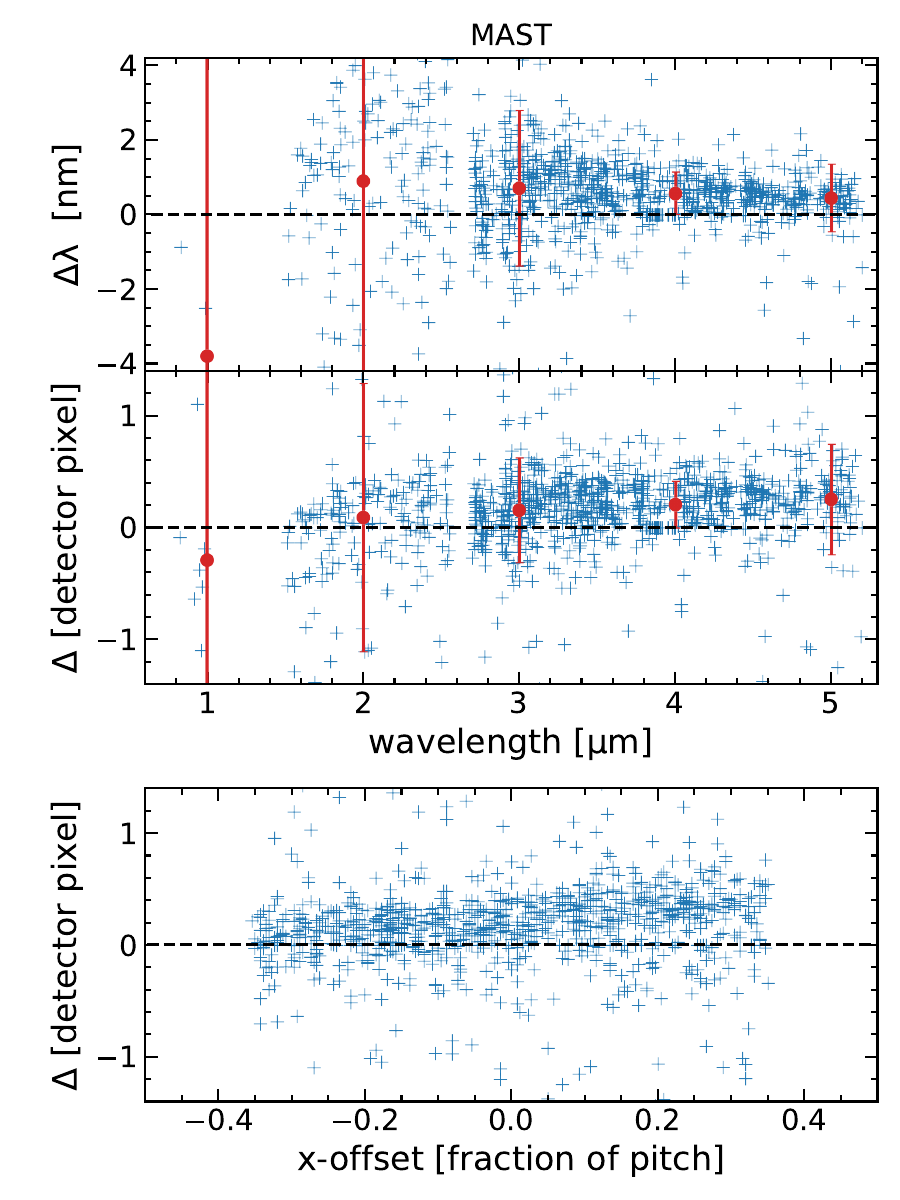}
\caption{Top and middle panels show the wavelength offset in wavelength and detector pixel units between PRISM and R1000 MAST spectra as a function of wavelength. Red marks report the median and the standard deviation values of the distribution of measurements at 1, 2, 3, 4, and 5$\mu$m. The bottom panel illustrates the wavelength offset in detector pixel units as a function of the intra-shutter location of the targets.}\label{f.app_mastwavecal}
\end{figure}

In this appendix, we show the wavelength discrepancy between the prism and grating spectra delivered by the MAST archive. We calculate the wavelength offset by measuring the line centroid of the brightest rest frame optical line, such as \OIII, \Halpha, and \HeI. The results are shown in  Figure~\ref{f.app_mastwavecal} We find a systematic offset of about 0.5 nm between the line centroids measured with the prism and those obtained with the grating. This value is consistent with the offset reported in DR3 (see Section~\ref{s.wavecal}). In addition, the spectral offset depends on the intra-shutter position of the targets, suggesting that the wavelength calibration step applied by the STScI pipeline does not provide an adequate correction.

\bibliographystyle{mnras}
\bibliography{config/astro} 

\end{document}